# Pedophysics: an open-source python package for soil geophysics.

**Authors:** Gaston Mendoza Veirana[1] (*Gaston.MendozaVeirana@ugent.be*), Philippe De Smedt[1,2] (*Philippe.DeSmedt@UGent.be*), Jeroen Verhegge[1,2] (*Jeroen.Verhegge@UGent.be*), Wim Cornelis[1] (*Wim.Cornelis@UGent.be*)

[1]*Department of Environment, Faculty of Bioscience Engineering, Ghent University, Coupure Links 653, geb. B, 9000 Ghent, Belgium.*

[2] *Department of Archaeology, Ghent University, Sint-Pietersnieuwstraat 35-UFO, 9000 Ghent, Belgium.*

## Abstract

This study introduces *Pedophysics*, an open-source Python package designed to facilitate solutions for users who work in the field of soil assessment using near-surface geophysical electromagnetic techniques. At the core of this software is the ability to translate geophysical data into specific soil properties (and vice-versa) using pedophysical models (PM). Pedophysical modelling techniques offer valuable insights into various realms including precision agriculture, soil health, resource prospecting, nutrient and land management, hydrogeology, and heritage conservation.

In developing a tool for pedophysical modelling, some challenges emerged: selecting suitable PMs from the extensive literature, adapting these to specific conditions, and ensuring adequate data availability. While addressing these, we designed an automated workflow that implements robust PMs (selected after a throughout review), apply different modelling approaches based on soil characteristics and targeted properties, and employs pedotransfer functions and assumptions to integrate missing soil data into PMs.

The capabilities of *Pedophysics* extend to handling complex scenarios such as fusing data from different instruments, incorporating continuous monitoring measurements, and soil calibration data. With these solutions, *Pedophysics* automates the process of deriving targeted soil and geophysical properties with state-of-art accuracy. Hereby, users can rely on *Pedophysics* to implement specific knowledge about pedophysical modeling. The software promotes global access to advanced soil geophysical solutions by being open-source and encouraging community contributions.

*Pedophysics* is written in pure Python and has minimal dependencies. It can be easily installed from the Python Package Index (PyPI).

## 1.0 Introduction

Near-surface geophysical electromagnetic (EM) techniques are proven tools to support soil ecosystem services. Their greatest potential lies in the ability to provide quick and high-resolution characterization of soils, capturing its spatial and temporal variability. The contribution of such techniques to soil assessment is extensive, encompassing various areas such as precision agriculture (Romero-Ruiz et al., 2018), soil health (Tabbagh et al., 2023), resources prospecting, nutrient and land management (Bennett et al., 2000), hydrogeology, and heritage conservation (Linford, 2006). Most used EM geophysical techniques include electromagnetic induction (EMI) survey, direct-current surveys such as electrical resistive tomography (ERT) and vertical electrical sounding (VES), ground penetrating radar (GPR) surveys, and continuous soil monitoring using time-domain refractometry (TDR), time domain transmissions (TDT), as well as impedance (I) and capacitance sensors (C). These techniques provide with geophysical properties data of electrical conductivity ($\sigma$, or resistivity ($\rho$)), dielectric permittivity ($\varepsilon$), and magnetic susceptibility ($\mu$). While basic interpretations can be directly derived from the collected data, specific soil applications require to characterize quantitatively some soil properties (and state variables) such as soil texture, bulk density, cation exchange capacity ($CEC$), or most importantly, salinity and water content (Corwin and Plant, 2005). Soil salinity is a measure of total salt concentrations in soil liquid, and significantly impacts agriculture by reducing crop yields, plant nutritional imbalances, and changes in soil tilth and permeability (Corwin and Yemoto, 2020). Moreover, the water content is a preferred target because of its central role in soil-plant interaction, groundwater assessment, soil ecological functioning, and climate, as well as its large influence on $\sigma$ and $\varepsilon$.

The models that link soil to geophysical properties (and vice-versa) are called pedophysical models (PM), which can be integrated into interpretation schemes (e.g., after inversion, or through incorporating this into forward modelling procedures). This modelling step, translating geophysical properties into soil properties (and vice versa), thus constitutes a key aspect of near surface exploration. Despite the diversity in applications of soil geophysical EM prospection (as in soil health and hydrogeophysics), pedophysical modeling is implemented through similar models and procedures (Revil et al., 2012; Romero-Ruiz et al., 2018).

Commonly, relative apparent geophysical information (this is, out-of-sensor data) is translated to quantify a target soil property by developing a local stochastic model, which is fitted to calibration data (Corwin and Lesch, 2005). In this work, we focus on deterministic modelling, but incorporate fitting solutions for specific cases. Hereby, generalized PMs are used, rather than field-specific stochastic models, that work primarily with absolute (or 'true') geophysical data.

In designing a straightforward tool to implement PMs, we identify four key challenges. First, selecting an appropriate PM is complicated by numerous models available in the literature (e.g., Glover, 2015; Wunderlich et al., 2013; van Dam et al., 2005). Furthermore, determining which models are well-validated from a theoretical or practical point of view can be onerous. This is because the measurement techniques, soil conditions, and modelling procedures are not standardized, leading to contradictory results. Second, even well-validated PMs are developed to suit specific conditions, such as a particular range of soil textures or EM frequencies. Third, when a PM is selected, it's applicability is determined by the availability of sufficient data for populating that model. The absence of specific soil information, such as the $CEC$ or particle density, can render some PMs useless. Lastly, pedophysical modelling may be applied when data from different sources are used and combined (Ciampalini et al., 2015; Romero-Ruiz et al., 2018). Examples of such data fusion include merging data that has been obtained at different EM frequencies, or at different timepoints (monitoring), as well as data that includes the target property of interest (calibration data).

To address these challenges, we propose effective approaches. Firstly, we conduct a comprehensive yet concise review to identify robust models from the literature that have been proven accurate in a wide range of soils. Secondly, we carefully consider the characteristics of a given soil and the specific property being targeted, allowing us to select the most suitable PM to provide a reliable solution. Lastly, we employ appropriate pedotransfer functions (PTFs) to predict the missing soil properties and integrate them into the population of PMs.

In this work, our objective is to integrate these solutions into a versatile, open-source and user-friendly Python package called *Pedophysics*. This software aims to automate the process of obtaining accurate soil and geophysical information using PMs with state-of-art accuracy. Ultimately, the aim of *Pedophysics* is to contribute to the facilitation of worldwide access to soil geophysical solutions. This is achieved by providing open-source code and encouraging contributions for further development. The reader committed to give *Pedophysics* a quick try finds coding examples in Section 4.

*Pedophysics* enables contributions to a list of specific soil applications, which include: delineating productivity zones (Corwin and Plant, 2005), quantifying the geophysical response to changes in specific soil properties (Hanssens et al., 2019), aquifer characterization (Revil et al., 2012), archaeological survey planning and feature characterization (Verhegge et al., 2023), soil compaction simulation (Romero-Ruiz et al., 2018), soil salinity determination (Corwin and Lesch, 2005), fusion of multisource geophysical data including field or laboratory soil calibration (Ciampalini et al., 2015) and time-lapse monitoring.

## 2 Pedophysical modelling

This section discusses the most accurate and thoroughly validated PMs, while presenting the scenarios in which these can be applied to link soil and geophysical properties.

### 2.1 Single frequency dielectric permittivity

Soil $\varepsilon_b$ is a key geophysical property commonly obtained by using GPR for near surface high resolution mapping, and moisture sensors (TDR, I and C) for in-depth continuous monitoring data (Topp et al., 1996). While $\varepsilon_b$ depends on multiple soil properties as well as the deployed EM frequency (Chen and Or, 2006), it is primarily linked to soil water content. Dielectric permittivity pedophysical models (PPMs) link soil properties with $\varepsilon_b$, and are generally developed for different frequency ranges. One of the most useful PPMs was proposed by Lichtenecker and Rother (1931) (LR):

$$\varepsilon_b^\alpha = \theta \varepsilon_w^\alpha + (1 - \emptyset)\varepsilon_s^\alpha + (\emptyset - \theta)\varepsilon_a^\alpha.$$

*Equation 1*

Valid for a three-phase soil, with $\theta$ the volumetric water content and $\varepsilon_w$ the soil water phase real relative dielectric permittivity, $\varepsilon_s$ the soil solid phase real relative dielectric permittivity, $\emptyset$ the soil porosity, $\varepsilon_a$ the soil air phase real relative dielectric permittivity, and $\alpha$ the alpha geometrical parameter. While $\varepsilon_a$ and $\varepsilon_s$ can be set as constants (1.2 and 4, respectively), $\varepsilon_w$ depends on temperature and salinity (Malmberg and Maryott, 1956; Revil et al., 1999) but it is normally set as 80. Soil porosity is calculated from bulk density ($b_d$) and particle density ($p_d$) using $\emptyset = 1 - b_d/p_d$, and $p_d$ can be calculated using a PTF (Schjønning et al., 2017).

The LR PPM (Equation 1) has shown accuracy across various soils measured with different instruments (Roth et al., 1990; Wunderlich et al., 2013). An additional advantage of this PPM is that EM frequency and soil texture dependance can be accounted for trough $\alpha$ (Mendoza Veirana et al., 2023). While in most cases $\alpha$ is fixed as 0.5, this parameter was found to be correlated positively with $CEC$ using data of an impedance moisture sensor collected at 50 MHz:

$$\alpha = 0.271 \log\left(CEC \frac{100g}{meq}\right) + 0.306,$$

*Equation 2*

which leads to more accurate soil $\varepsilon_b$ modelling. For data collected with GPR (at 1.6 GHz) Wunderlich et al. (2013) found a negative correlation with clay content:

$$\alpha = -0.46 clay + 0.71$$

*Equation 3*

Lastly, the frequency range between 100 MHz to 200 MHz remains more uncertain than other frequency ranges, because of the presence of the cross-over frequency (Chen and Or, 2006), for which a conservative parameter setting ($\alpha = 0.5$) seems more effective.

## 2.2 Direct current electrical conductivity

Electrical conductivity pedophysical models (ECPMs) have traditionally been developed for consolidated sands, as the case of the Archie equations (Archie, 1942). More accurate ECPMs consider finer soil fractions (through their surface conductivity) such as the model developed by (Waxman and Smits, 1968):

$$\sigma_b = \sigma_w S^n \emptyset^m + S^{n-1} \emptyset^m \sigma_{surf}.$$

*Equation 4*

Here, $\sigma_w$ is the water EC and $\sigma_{surf}$ is the EC of solid surfaces, which arises due to clay particles and the ionic double layer that envelops them. Although the term $\sigma_{surf}$ depends on porosity and $CEC$, it must be calibrated against empirical data, as well as the geometrical parameters $m$ and $n$. Another widely tested ECPM, involving similar soil properties, was proposed by Rhoades et al. (1976):

$$\sigma_b = \sigma_w(E\theta^2 + F\theta) + \sigma_s$$

*Equation 5*

In this equation, the empirical constants $E$ and $F$ are determined by fitting the model to calibration data, and $\sigma_s$ is the bulk surface conductivity.

As more extensive datasets have become available over recent decades, empirical expressions for geometric parameters become available. For instance, a precise empirical formula for $\sigma_{surf}$ was developed by Doussan and Ruy (2009):

$$\sigma_{surf} = 0.654 \frac{clay}{sand + silt} + 0.018; \; R^2 = 0.97$$

*Equation 6*

Recently, Fu et al. (2022) introduced a robust ECPM by combining Equation 4 with Glover et al. (2000) ECPM resulting in an expression similar to Equation 5:

$$\sigma_b = \sigma_w \theta^w + \sigma_{surf} \theta^{w-1} \emptyset + \sigma_s (1 - \emptyset)^s$$

*Equation 7*

where $w$ and $s$ are constants with average values of 2 and 1, respectively. This calibration was carried out using data from 15 soil samples collected by various authors (Rhoades et al., 1976), with clay contents ranging from 0 to 33%, and bulk densities from 1.05 to 1.83 $g/cm^3$. Then, Equation 7 becomes:

$$\sigma_b = \sigma_w \theta^2 + \theta \emptyset \left( 0.654 \frac{clay}{100 - clay} + 0.018 \right) + (1 - \emptyset)\sigma_s; \; R^2 = 0.98$$

*Equation 8*

The term $sand + silt$ was replaced with $100 - clay$ in Equation 6 to minimize the soil information requirements. This ECPM was validated with independent soil data from six samples taken from six different studies, consisting of various salinities and water contents, amounting to around 300 data points. It achieved an unprecedented $R^2$ value of 0.98 (Fu et al., 2022). Additionally, this model relies on easily obtainable soil properties, and allows for the integration of $\sigma_{sat}$ and $\sigma_{dry}$ data (if available) as follows:

$$\sigma_w = \frac{\sigma_{sat} - \sigma_{dry}}{\emptyset^2} - \sigma_{surf}$$

*Equation 9*

To the best of our knowledge, Fu's model (Equation 8) is the most thoroughly validated non-fitting ECPM and is thus implemented in *Pedophysics*.

## 2.3 Soil salinity

Soil salinity is normally obtained through its link with $\sigma_w$, being the model of (Sen and Goode, 1992a, b) the most used empirical formula for a NaCl aqueous solution:

$$\sigma_w = (d_1 + d_2 T + d_3 T^2) C_f - \left(\frac{d_4 + d_5 T}{1 + d_6 C_f^{0.5}}\right) C_f^{1.5}$$

*Equation 10*

Where $d_1 = 5.6\ Sl/mol\ m$, $d_2 = 0.27\ Sl/mol\ m\ °C$, $d_3 = -1.51 * 10^{-4}\ Sl/mol\ m\ °C^2$, $d_4 = 2.36\ S/m * (mol/l)^{-1.5}$, $d_5 = 0.099\ S/m°C * (mol/l)^{-1.5}$, $d_6 = 0.214\ (mol/l)^{-1.5}$, $T$ is the temperature of the fluid in °C, and $C_f$ is the salinity of the pore fluid ($mol/l$). As most geophysical techniques focus to characterize soil salinity through $\sigma_w$, we discuss adequate ECPMs to calculate $\sigma_w$.

Approaches to obtain $\sigma_w$ from EM soil data can be divided into non-fitting and fitting modeling procedures. The non-fitting approach uses a ECPM (as Equation 8) when $\sigma_b$ and ancillary soil properties are known (i.e., $\theta, \emptyset, clay$).

The fitting approach, on the other hand uses two kinds of PMs: a linear $\varepsilon_b$-$\sigma_b$ PM (Hilhorst, 2000; Malicki and Walczak, 1999), and a $\sigma_b$-$\theta^2$ PM (as Equation 5) (Hamed et al., 2003; Rhoades et al., 1976). Such an approach is usually more precise that the non-fitting counterpart, since it uses calibration data to render a single $\sigma_w$ value.

The linear $\varepsilon_b$-$\sigma_b$ modelling is relatively precise and easy to implement, this is why most moisture sensors (TDR, C, I) provide $\sigma_w$ using this technique. This consists of fitting a linear model between $\varepsilon_b$ and $\sigma_b$, while $\sigma_w$ is obtained from the slope ($A$) of the line:

$$\varepsilon_b = \sigma_b A + B$$

*Equation 11*

where $A$ and B are empirical constants. The most relevant expression for $A$ and B are proposed by (Hilhorst, 2000):

$$\varepsilon_b = \sigma_b \frac{\varepsilon_w}{\sigma_w} + \varepsilon_{offset},$$

*Equation 12*

where $\varepsilon_{offset}$ is the intercept of the $\varepsilon_b$-$\sigma_b$ fitted line. This method has shown relatively good results in certain conditions, but the evidence across the literature is contradictory (See e.g., Bañón et al., 2021; Brovelli, A. & G. Cassiani, 2011).

Calculations of $\sigma_w$ based on $\sigma_b$-$\theta^2$ fitting PM are validated for a wide range of soil types, varying water contents and salinities (Amente et al., 2000; Corwin and Yemoto, 2020; Hamed et al., 2003). Most accurate results were found using the PM proposed by Rhoades et al. (1976) (Equation 5).

The prediction capacity of Equation 5 and Equation 12 was evaluated by (Hamed et al., 2003), using 9 soil samples of various soil textures, $\sigma_w$, and $\theta$. It was concluded that $A$ and B are not constant but dependent on soil type and $\sigma_w$ (confirming results from Malicki & Walczak, 1999; and Persson, 2002).

Overall, Equation 11 produced poor estimations of $\sigma_w$ (RMSE from 0.03 to 0.1 S/m) compared to estimations of $\sigma_w$ obtained by fitting of Equation 5 (RMSE from 0.015 to 0.048 S/m), especially in saline dry sands. Furthermore, while Equation 5 is applicable to DC frequencies, the influence of frequency modulations on the accuracy of Equation 12 has not been evaluated.

**2.4 Incorporating calibration data through fitting PMs.**

While most of implementations of PMs in soil geophysics integrate calibration data through fitting approaches, few studies compare the performance of different PMs that can be deployed. Using soil data obtained at 1.6 GHz, Wunderlich et al. (2013) conclude that two effective medium PMs (Equation 13 and Equation 14) perform best for estimating $\varepsilon_b$ and $\sigma_b$ in direct-current (DC) regime, respectively:

$$\frac{d\varepsilon(p)}{dp} = \frac{\varepsilon(p)(\theta_n - \theta_1)}{1 + \theta_1 - \theta_n + p(\theta_n - \theta_1)} \frac{\varepsilon_w - \varepsilon(p)}{L_w \varepsilon_w + (1 - L_w)\varepsilon(p)}$$

$\varepsilon(p = 0) = \varepsilon_1;\ \varepsilon(p = 1) = \varepsilon_n$

*Equation 13*

$$\frac{d\sigma(p)}{dp} = \frac{\sigma(p)(\theta_n - \theta_1)}{1 + \theta_1 - \theta_n + p(\theta_n - \theta_1)} \frac{\sigma_w - \sigma(p)}{L_w \sigma_w + (1 - L_w)\sigma(p)}$$

$\sigma(p = 0) = \sigma_1;\ \sigma(p = 1) = \sigma_n$

*Equation 14*

The geometrical parameter $L_w$ is the depolarization factor of the soil water phase, normally optimized for fitting. The integration variable $p$ ranges from 0 to 1, $\varepsilon(p = 0) = \varepsilon_1$ is the initial $\varepsilon_b$ that corresponds to the initial volumetric water content ($\theta_1$), and $\varepsilon(p = 1) = \varepsilon_b$ corresponds to $\theta$, similarly for $\sigma$ in Equation 14. For estimating $\varepsilon_b$, Equation 13 was further validated by Mendoza Veirana et al. (2023), where several fitting PPMs were tested using data obtained at 50 MHz of 10 soils with varying texture.

**2.5 Soil dielectric dispersion**

The dielectric dispersion in soils refers to how their observed dielectric properties vary with the frequency of the applied EM field (González-Teruel et al., 2020). Dielectric dispersion PMs are used when geophysical data is obtained at different EM frequencies. This is particularly relevant when multi-sensor data are used, such as in dual-frequency surveys, joint inversions, vector analyzer experiments, or when assessing the likelihood of detecting specific soil features for surveys at different EM frequencies.

Generally, the frequency dependence of $\sigma_b$ increases with EM frequency ($f_\sigma$) (Alipio and Visacro, 2014). An ECPM for dielectric dispersion combines two types of dispersion: a constant one, referred to as the conventional DC $\sigma_b$ ($\sigma_{dc}$), associated with free ions and electrons, and a frequency-dependent one linked to losses caused by polarization processes (Zhou et al., 2015).

The frequency dependence of $\varepsilon_b$ is attributed to the different dielectric relaxation responses of the soil phases and their interactions (Dobson et al., 1985; González-Teruel et al., 2020). This process is strongly influenced by the Maxwell-Wagner effect at EM frequencies approximately <100 MHz (Chen and Or, 2006; Schwartz et al., 2009). Within this range, all factors influencing $\sigma_b$ are positively correlated to $\varepsilon_b$ (e.g., clay content, salinity, temperature). For higher frequencies, the Maxwell-Wagner effect becomes increasingly negligible and $\varepsilon_b$ depends more exclusively on water content. At such higher frequencies, increases in other soil properties such as salinity or clay led to a decreasing $\varepsilon_b$. Therefore, a suitable dielectric dispersion PM should not only account for soil texture, salinity, temperature, Maxwell-Wagner effect (in the case of $\varepsilon_b$), soil phases (at least three), but also consider a wide range of commonly used electromagnetic frequencies.

Normally, dielectric dispersion PMs do not meet all these requirements (see e.g., the PMs presented by: (Alipio and Visacro, 2014; Chen and Or, 2006; Dobson et al., 1985; González-Teruel et al., 2020;

Hallikainen et al., 1985). In conclusion, as (González-Teruel et al., 2020)) stated, there is still a need to develop a general three-phase MHz–GHz dielectric PM.

Despite several dielectric dispersion PMs being available in literature, few comparative studies have been conducted (e.g., (Alipio and Visacro, 2014; Cavka et al., 2014; van Dam et al., 2005). An empirical and theoretical test of six semiempirical models was presented by (Cavka et al., 2014). Models were hereby compared using 18 partially saturated sandy soil samples with varying $\sigma_w$ (data from Bigelow and Eberle, 1983; He et al., 2013). The study concluded that the semiempirical model proposed by (Longmire and Smith, 1975) (LS) is among the best performing. Additionally, the LS model is defined for the largest frequency range between other models (5 Hz to 30 GHz, see Longmire and Smith, 1975) and the fitting of parameters is not needed:

$$\varepsilon_b(f_\varepsilon) = \varepsilon_{inf} + \sum_{i=1}^{13} \frac{a_i}{1+\left(f_\varepsilon/F_i\right)^2},$$

*Equation 15*

$$\sigma_b(f_\sigma) = \sigma_{dc} + 2\pi\varepsilon_0 \sum_{i=1}^{13} a_i F_i \frac{\left(f_\sigma/F_i\right)^2}{1+\left(f_\sigma/F_i\right)^2},$$

*Equation 16*

$$F_i = (125\sigma_{dc})^{0.8312} * 10^{i-1},$$

*Equation 17*

where $a_i$ are empirical constants, $\varepsilon_{inf}$ is the soil bulk real relative permittivity at infinite frequency, and $f_\sigma$ is the frequency of the electric conductivity measurement.

However, a limitation remains in the LS model (Equation 15, Equation 16, and Equation 17), as it was developed based on single-texture soil data, and the Maxwell-Wagner effect does not seem to be accounted for at $f_\varepsilon$>100 MHz. Based on the calibration and tested samples using the LS model, it is likely that this works well for low relaxation soils, but it is not suitable for wet clays where the relaxation is large (Jones et al., 2005). This limitation leads to similar $\varepsilon_b$ predictions for wet sands and dry clays with similar $\sigma_b$, while their response is expected to be dissimilar (González-Teruel et al., 2020; Jones et al., 2005).

**2.6 Temperature correction**

Soil temperature affects soil EM geophysical properties, and its effect should generally be considered to ensure optimal soil modelling. The effect of $T$ on $\varepsilon_b$, for instance, varies with frequency. However, as the $T$ - $\varepsilon_b$ relationship has not been fully explained with current PPMs, added to its relatively low effect (Chen and Or, 2006b), we do not delve into this. However, the impact of $T$ in $\sigma_b$ is more significant and well understood. Several temperature correction ECPMs were compared by Ma et al. (2010) in the range of approximately 0 to 50 °C, highlighting the accuracy of the model proposed by Sheets and Hendrickx (1995):

$$\sigma_{tc} = \sigma_b \left(0.447 + 1.4034 e^{-T+273.15/26.815}\right)$$

*Equation 18*

Where $\sigma_{tc}$ is the soil bulk real electrical conductivity temperature corrected, this is, the $\sigma_b$ at 298.15 K (25 °C). When $\sigma_{dc}$ is corrected by temperature, we obtain $\sigma_{dc\ tc}$.

## 3 *Pedophysics*: functioning and structure

This section synthesizes the concepts of pedophysical modelling (Section 2) with their implementation in the *Pedophysics* Python package.

Figure 1 presents a general scheme of *Pedophysics*´ usage. Initially, the user provides soil data and specifies a target property. *Pedophysics* then fills in any missing data, using pedotransfer functions and assumptions. Next, the target property is calculated using the most suitable approach.

Afterwards, if the calculated target property contains incomplete elements (indicated by NaNs, i.e. not a number), additional data must be supplied to ensure a complete target prediction.

*Pedophysics* architecture is structured according to the workflow described earlier, as depicted in Figure 2 Pedophysics directory, containing the subpackages pedophysical_models, pedotransfer_functions, predict, and utils. Modules of the main package are __init__.py, instruments.py, main.py, and simulate.py.. It comprises four subpackages and four modules. The subpackages *pedophysical_models* and *pedotransfer_functions* include modules that implement the PMs and PTFs detailed in Section 2. These functions are then utilized in the subpackage *predict* (see Section 3.2). Additionally, the *utils* subpackage provides a set of statistical tools. Lastly, within the main package the *simulate.py* module is responsible for defining virtual soil samples.

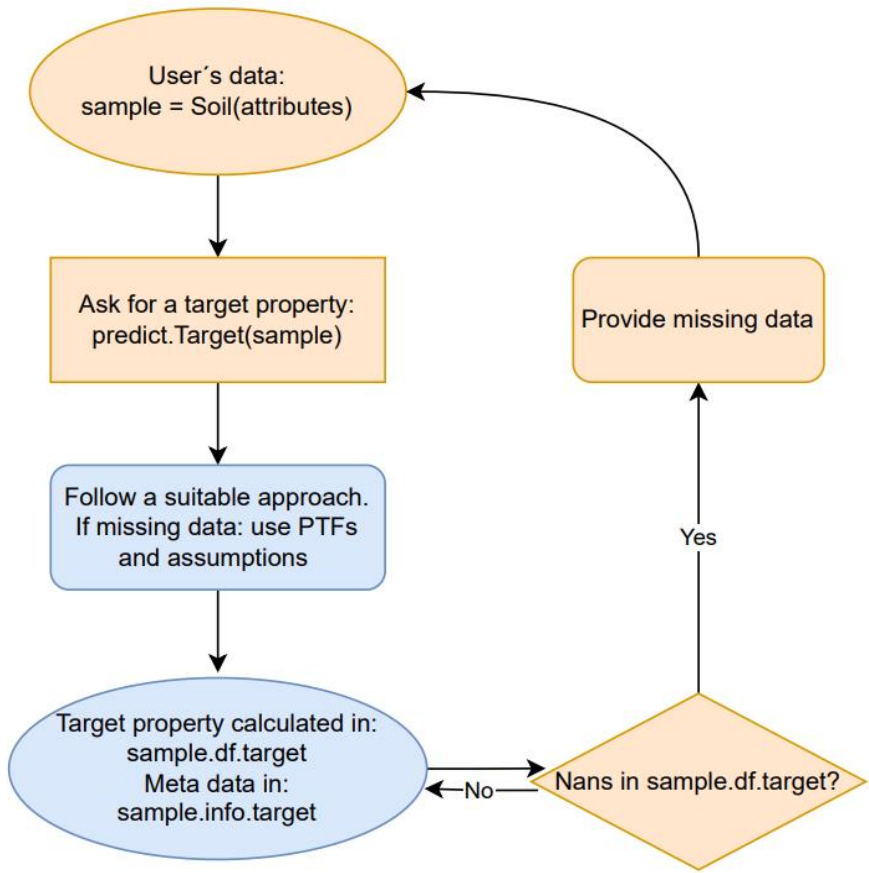

*Figure 1 Workflow of* Pedophysics´ *use. Yellow icons correspond to actions done by the user, blue icons for automated actions by* Pedophysics.

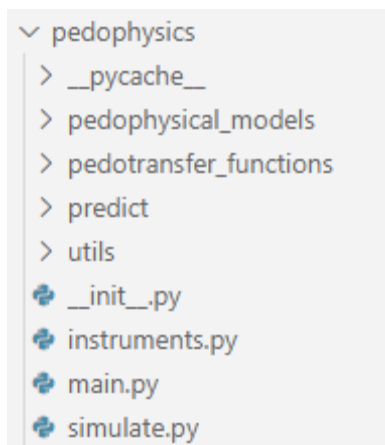

*Figure 2 Pedophysics directory, containing the subpackages pedophysical_models, pedotransfer_functions, predict, and utils. Modules of the main package are __init__.py, instruments.py, main.py, and simulate.py.*

### 3.1 Defining a virtual soil (*pedophysics.simulate.py*)

The core of *pedophysics* is the *simulate.py* module, where soils are defined through a Python class named *Soil*. Consequently, users can create a soil sample instance using the *Soil* class by specifying its attributes, which represent various soil and geophysical properties, alongside model factors (listed in Table 1).

| Attribute | Symbol | Description | Units | Type | Set |
|---|---|---|---|---|---|
| **temperature** | $T$ | Soil bulk temperature | [K] | float, np.float64, int, np.ndarray, list | 298.15 |
| **water** | $\theta$ | Soil volumetric water content | [m**3/m**3] | float, np.float64, int, np.ndarray, list | No |
| **salinity** | $C_f$ | Soil salinity (NaCl) of the bulk pore fluid | [mol/L] | float, np.float64, int, np.ndarray, list | No |
| **sand** | $sand$ | Soil sand content | [g/g]*100 | float, np.float64, int, np.ndarray, list | No |
| **silt** | $silt$ | Soil silt content | [g/g]*100 | float, np.float64, int, np.ndarray, list | No |
| **clay** | $clay$ | Soil clay content | [g/g]*100 | float, np.float64, int, np.ndarray, list | No |
| **porosity** | $\emptyset$ | Soil porosity | [m**3/m**3] | float, np.float64, int, np.ndarray, list | No |
| **bulk_density** | $b_d$ | Soil bulk density | [kg/m**3] | float, np.float64, int, np.ndarray, list | No |
| **particle_density** | $p_d$ | Soil particle density | [kg/m**3] | float, np.float64, int, np.ndarray, list | 2.65 |
| **CEC** | $CEC$ | Soil cation exchange capacity | [meq/100g] | float, np.float64, int, np.ndarray, list | No |
| **orgm** | $orgm$ | Soil organic matter content | [g/g]*100 | float, np.float64, int, np.ndarray, list | No |
| **bulk_perm** | $\varepsilon_b$ | Soil bulk real relative dielectric permittivity | [-] | float, np.float64, int, np.ndarray, list | No |
| **bulk_perm_inf** | $\varepsilon_{inf}$ | Soil bulk real relative permittivity at infinite frequency | [-] | float, np.float64, int, np.ndarray, list | 5 |
| **water_perm** | $\varepsilon_w$ | Soil water phase real dielectric permittivity | [-] | float, np.float64, int, np.ndarray, list | 80 |
| **solid_perm** | $\varepsilon_s$ | Soil solid real relative dielectric permittivity phase | [-] | float, np.float64, int, np.ndarray, list | 4 |
| **air_perm** | $\varepsilon_a$ | Soil air real relative dielectric permittivity phase | [-] | float, np.float64, int, np.ndarray, list | 1.2 |
| offset_perm | $\varepsilon_{offset}$ | Soil bulk real relative dielectric permittivity when soil bulk real electrical conductivity is zero | [-] | float, np.float64, int, np.ndarray, list | No |
| bulk_ec | $\sigma_b$ | Soil bulk real electrical conductivity | [S/m] | float, np.float64, int, np.ndarray, list | No |
| bulk_ec_tc | $\sigma_{tc}$ | Soil bulk real electrical conductivity temperature corrected (298.15 K) | [S/m] | float, np.float64, int, np.ndarray, list | No |
| bulk_ec_dc | $\sigma_{dc}$ | Soil bulk real electrical conductivity direct current (0 Hz) | [S/m] | float, np.float64, int, np.ndarray, list | No |
| bulk_ec_dc_tc | $\sigma_{dc\,tc}$ | Soil bulk real electrical conductivity direct current (0 Hz) temperature corrected (298.15 K) | [S/m] | float, np.float64, int, np.ndarray, list | No |
| bulk_ec_tc | $\sigma_{tc}$ | Soil bulk real electrical conductivity temperature corrected | [S/m] | float, np.float64, int, np.ndarray, list | No |
| water_ec | $\sigma_w$ | Soil water real electrical conductivity | [S/m] | float, np.float64, int, np.ndarray, list | No |
| s_ec | $\sigma_s$ | Soil bulk real surface electrical conductivity | [S/m] | float, np.float64, int, np.ndarray, list | No |
| solid_ec | $\sigma_{solid}$ | Soil solid real electrical conductivity | [S/m] | float, np.float64, int, np.ndarray, list | No |
| dry_ec | $\sigma_{dry}$ | Soil bulk real electrical conductivity at zero water content | [S/m] | float, np.float64, int, np.ndarray, list | No |
| sat_ec | $\sigma_{sat}$ | Soil bulk real electrical conductivity at saturation water content | [S/m] | float, np.float64, int, np.ndarray, list | No |
| frequency_perm | $f_\varepsilon$ | Bandwidth centroid frequency of dielectric permittivity measurement | [Hz] | float, np.float64, int, np.ndarray, list | No |
| frequency_ec | $f_\sigma$ | Bandwidth centroid frequency of electric conductivity measurement | [Hz] | float, np.float64, int, np.ndarray, list | 0 |
| L | $L$ | Soil scalar depolarization factor of solid particles (effective medium theory) | [-] | float, int, np.float64 | No |
| Lw | $L_w$ | Soil scalar depolarization factor of water aggregates (effective medium theory) | [-] | float, int, np.float64 | No |
| m | $m$ | Soil cementation factor as defined in Archie law | [-] | float, int, np.float64 | No |
| n | $n$ | Soil saturation factor as defined in Archie second law | [-] | float, int, np.float64 | No |
| alpha | $\alpha$ | Soil alpha exponent as defined in volumetric mixing theory | [-] | float, int, np.float64 | No |
| E | $E$ | Empirical constant as in Rohades model | [-] | float, int, np.float64 | No |
| F | $F$ | Empirical constant as in Rohades model | [-] | float, int, np.float64 | No |
| texture | | Soil texture according to USDA convention: "Sand", "Loamy sand", "Sandy loam", "Loam", "Silt loam", "Silt", "Sandy clay loam", "Clay loam", "Silty clay loam", "Sandy clay", "Clay", "Silty clay" | [-] | str, numpy.nan | No |
| instrument | | Instrument utilized: 'HydraProbe', 'TDR', 'GPR', 'Miller 400D', 'Dualem' | [-] | str, numpy.nan | No |
| info | | Data Frame containing descriptive information about how each array-like attribute was calculated. | [-] | pandas.DataFrame | No |

| | | | | |
|---|---|---|---|---|
| **df** | Data Frame containing the quantitative information of all soil array-like attributes for each state. | [-] | pandas.DataFrame | No |
| **roundn** | Number of decimal places to round results. | [-] | int | 3 |
| **range_ratio** | Ratio to extend the domain of the regression by fitting approach. | [-] | float, int, np.float64 | 2 |
| **n_states** | Number of soil states. | [-] | float, int, np.float64 | No |

*Table 1. Soil Class attributes that represent soil and geophysical properties, and model factors. Rows in blue list array-like attributes, withe list single-value attributes, and yellow list special attributes.*

The *Soil* attributes are divided into three groups: **single-value** (e.g., $L_w$, $m$, texture class), **array-like** (e.g., $\theta$, $C_f$, $\sigma_{dc}$, and $\varepsilon_b$), and **special** attributes. The difference between the first two is that single-value attributes are defined just as Int, Float, and np.float64 object types, while array-like attributes can also be defined as a list or NumPy array type. Attributes defined with a wrong type are warned with an exception (ValueError). Array-like attributes are used to simulate the evolution of such variable through time or space. For example, to simulate a soil of which the water content changes over time and which is recorded at six different moments, we define *Soil.water* as a list (or NumPy array) with six elements (see Figure 3. Definition of a virtual soil as a Soil instance. sample1 is a Soil instance with two given attributes: water and porosity. water is an array-like attribute composed by six soil states, and porosity is an array-like attribute defined as a constant. The array-like attribute porosity of sample1 is automatically converted to a NumPy.Array with length equal to n_states and filled with the value of the state zero (0.4). The attribute water_ec was not defined by the user, then it is set to length equal to n_states and filled with Numpy.nan.). Thus, we refer as soil state to the element number of an array-like *Soil* attribute. A special *Soil* attribute is *n_states*, defined as the maximum length of all array-like attributes. All the attributes that are not defined by the user are assumed as NaN (always represented by Numpy.nan).

Two automatic modifications are applied to array-like attributes. First, these are all converted to the numpy.ndarray type, and their lengths are standardized to *n_states*, by populating the missing states with Numpy.nan. Secondly, if only the state number zero is defined by the user for an array-like attribute, this value is treated as a constant for all states from zero to *n_states* (see Figure 3).

**Input:**

```python
from pedophysics.simulate import Soil

sample1 = Soil(water = [0.1, 0.15, 0.2, 0.25, 0.3, 0.4],
               porosity = 0.4)

sample1.water
sample1.porosity
sample1.water_ec
```

**Output:**

```
>>>array([0.1, 0.15, 0.2, 0.25, 0.3, 0.4])
>>>array([0.4, 0.4, 0.4, 0.4, 0.4, 0.4])
>>>array([nan, nan, nan, nan, nan, nan])
```

*Figure 3. Definition of a virtual soil as a Soil instance. sample1 is a Soil instance with two given attributes: water and porosity. water is an array-like attribute composed by six soil states, and porosity is an array-like attribute defined as a constant. The array-like attribute porosity of sample1 is automatically converted to a NumPy.Array with length equal to n_states and filled with the value of the state zero (0.4). The attribute water_ec was not defined by the user, then it is set to length equal to n_states and filled with Numpy.nan.*

An important special *Soil* attribute named *df* is defined as a pandas.dataframe object. It contains all array-like attributes within its columns, with each row representing a different soil state. All modifications to *Soil* attributes through the *predict* subpackage are recorded in *Soil.df*, while the *Soil* instance remains unmodified. Printing a *Soil* instance gives a print of *Soil.df* (see Figure 4).

Lastly, another special *Soil* attribute is *info*, a pandas.dataframe object. It holds descriptive information on how each array-like attribute is determined or modified by the *predict* subpackage at each state. This includes details about the utilized PM, the directory path of the function, and a reported or calculated error indicator. The reported accuracy scores of all PMs and PTFs are provided (when available), when fitting approaches are followed, the reached accuracy is provided.

**Input:**

```
sample1 = Soil(water = [0.1, 0.15, 0.2, 0.25, 0.3, 0.4],
               porosity = 0.4)
sample1.df
```

**Output:**

|   | temperature | water | salinity | sand | silt | clay | porosity | bulk_density | particle_density | CEC | ... | bulk_ec | bulk |
|---|---|---|---|---|---|---|---|---|---|---|---|---|---|
| 0 | NaN | 0.10 | NaN | NaN | NaN | NaN | 0.4 | NaN | NaN | NaN | ... | NaN |
| 1 | NaN | 0.15 | NaN | NaN | NaN | NaN | 0.4 | NaN | NaN | NaN | ... | NaN |
| 2 | NaN | 0.20 | NaN | NaN | NaN | NaN | 0.4 | NaN | NaN | NaN | ... | NaN |
| 3 | NaN | 0.25 | NaN | NaN | NaN | NaN | 0.4 | NaN | NaN | NaN | ... | NaN |
| 4 | NaN | 0.30 | NaN | NaN | NaN | NaN | 0.4 | NaN | NaN | NaN | ... | NaN |
| 5 | NaN | 0.40 | NaN | NaN | NaN | NaN | 0.4 | NaN | NaN | NaN | ... | NaN |

6 rows × 27 columns

*Figure 4. df attribute stores all array-like attributes. The Pandas.DataFrame object Soil.df contains all the Soil array-like attributes on its columns, and states in its rows. Here water and porosity are as given in the input of sample1 (Figure 3).*

### 3.2 Obtaining soil and geophysical properties (*predict* subpackage)

The prediction of a specific soil property (the target attribute) is enabled by the *predict* subpackage. Comprising various modules, each one specializes in offering solutions for distinct attributes and is named accordingly (see Figure 5. Directory of the subpackage predict.). Within *predict*, all modules are structured around a main function and, where necessary, sub-functions. For instance, to obtain the attribute *bulk_ec* the user uses the function *predict.BulkEC* (located in the module *pedophysics.predict.bulk_ec.py*).

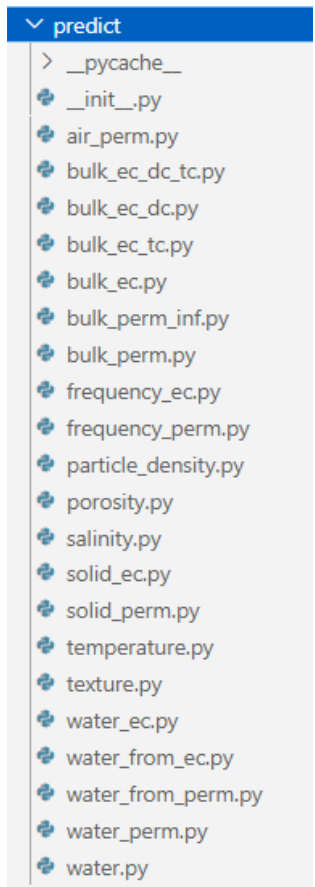

*Figure 5. Directory of the subpackage predict.*

The most relevant attributes that can be calculated using the *predict* module are *water*, *salinity*, *water_ec*, *bulk_ec*, and *bulk_perm.* To predict these attributes, soil data given by the user is automatically fused by hierarchy. A good example of hierarchical data fusion is the implementation of *non_fitting* and *fitting* functions, where the priority is always to obtain a solution by fitting. Alternatively, a solution is sought following non-fitting approaches, in which PMs are applied that do not require optimizing attributes, and the solution at each soil state is independent of the rest. In the software, *fitting* functions are deployed when the user provides calibration data (i.e., at least three soil states that match a specific condition), allowing to obtain a specific solution. These functions look for a best fit of the calibration data using a PM by optimizing some of its attributes. The domain in which the fitted function is valid is calculated using the special attribute *Soil.range_ratio*. When no calibration data is provided, or in case any state of the target property remains unknown after attempting *fitting* functions, , *non_fitting* functions are deployed.

### 3.2.1 Electromagnetic properties of soils (*predict.bulk_perm.py* and *predict.bulk_ec.py*)

Predicting *Soil.bulk_perm* and *Soil.bulk_ec* in *Pedophysics* is done following similar approaches where *Soil.water*, *Soil.frequency_ec*, and *Soil.frequency_perm* play a major role. The structure of sub functions follows a similar organization, but mayor differences remain on the modelling aspect.

Predicting *Soil.bulk_perm* in *Pedophysics* (see Figure 6. Pedophysics workflow for the calculation of $\varepsilon_b$. Each block describes a function, which name is at the heading. Light yellow heading is used for the main function predict.bulk_perm.BulkPerm, while light blue headings refer to secondary functions. Black horizontal lines divide a block into different conditions. Light red blocks refer to mandatory soil attributes for a condition. Purple blocks refer to mandatory soil attributes that can be derived from other ones.) is done following the methods described in Section 2.1, 2.4, and 2.5. A mandatory attribute is *Soil.frequency_perm*, if this is not given there is no possible solution and *Soil.bulk_perm* remains unmodified. If *Soil.frequency_perm* is given, functions for either constant or changing frequency are called. The last calculates the target using the *LongmireSmithP* function (Equation 15 and Equation 17), and the former case is divided into fitting and non-fitting approaches. When at least three soil states have *Soil.water* and *Soil.bulk_perm* data (i.e., calibration data), the function *fitting* uses *WunderlichP* function (Equation 13) by fitting its depolarization factor $L_w$ (*Soil.Lw*). The range of values of *Soil.water* where the fitted model works is defined in *water_range*. This range can be extended by giving a smaller *Soil.range_ratio* (equal to 2 by default). If any state of *Soil.bulk_perm* is still unknown, the *non_fitting* function applies different PMs in four different frequency ranges: 5 to 30e6 Hz uses *LongmireSmithP*, 30e6 to 100e6 Hz uses the function *LR_MV* (Equation 1 and Equation 2), 100e6 to 200e6 Hz uses *LR* (Equation 1 with alpha = 0.5), and 200e6 to 30e9 Hz uses *LR_W* (Equation 1 and Equation 3).

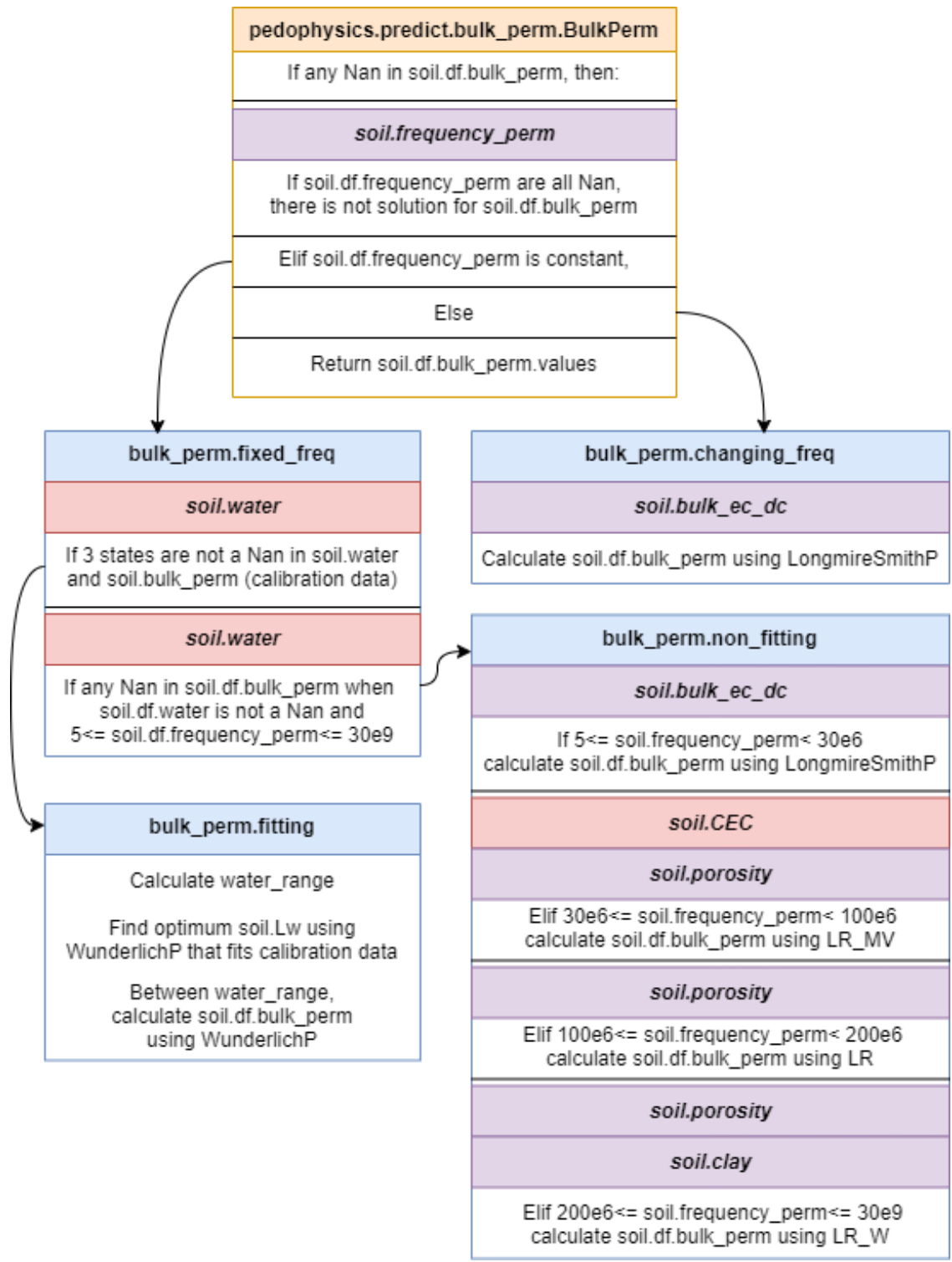

*Figure 6. Pedophysics workflow for the calculation of $\varepsilon_b$. Each block describes a function, which name is at the heading. Light yellow heading is used for the main function predict.bulk_perm.BulkPerm, while light blue headings refer to secondary functions. Black horizontal lines divide a block into different conditions. Light red blocks refer to mandatory soil attributes for a condition. Purple blocks refer to mandatory soil attributes that can be derived from other ones.*

Predictions of *Soil.bulk_ec* and *Soil.bulk_ec_dc* are based on calculating *Soil.bulk_ec_dc_tc* and applying temperature and frequency corrections, respectively.

Similarly, to predict *Soil.bulk_ec_dc_tc* (Figure 7 Pedophysics workflow for the calculation of $\sigma_{dc\ tc}$. Each block describes a function, which name is at the heading. Light yellow heading is used for the main function predict.bulk_ec_dc_tc.BulkECDCTC, while light blue headings refer to secondary functions. Black horizontal lines divide a block into different conditions. Light red blocks refer to mandatory soil attributes for a condition. Purple blocks refer to mandatory soil attributes that can be derived from other ones.) frequency and temperature corrections have place in the *shift_to_bulk_ec_dc_tc* function using Equation 16 and Equation 18, respectively. From a modelling perspective, *Pedophysics* assumes DC regime (i.e., *frequency_ec = 0*) if *frequency_ec* is not given (this is, Numpy.nan) or < 5 Hz. Also, the default $T$ is 298.15 K, therefore, if no *frequency_ec* and *temperature* is given, *Soil.bulk_ec* is equal to *Soil.bulk_ec_dc_tc*.

In case that calibration data in *Soil.water* and *Soil.bulk_ec_dc_tc* are provided, the *fitting* function uses the *WunderlichEC* function (Equation 14). If NaNs remain in *Soil.bulk_ec_dc_tc*, the *non_fitting* function uses *Fu* function (Equation 8).

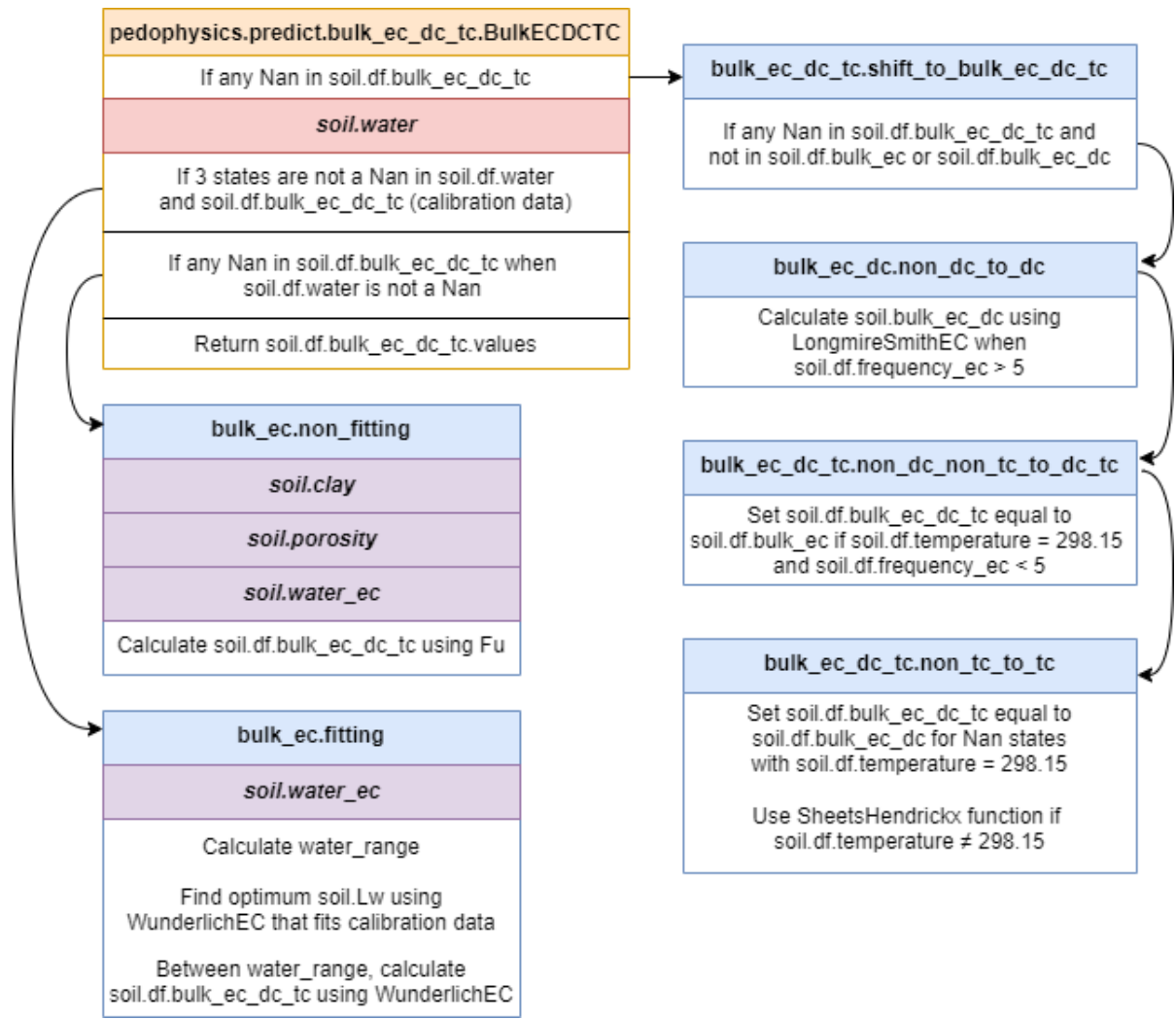

Figure 7 Pedophysics workflow for the calculation of $\sigma_{dc\,tc}$. Each block describes a function, which name is at the heading. Light yellow heading is used for the main function predict.bulk_ec_dc_tc.BulkECDCTC, while light blue headings refer to secondary functions. Black horizontal lines divide a block into different conditions. Light red blocks refer to mandatory soil attributes for a condition. Purple blocks refer to mandatory soil attributes that can be derived from other ones.

### 3.2.2 Water content prediction (*predict.water.py*)

The *predict.water.py* module enables predicting *Soil.water* based on *Soil.bulk_perm* and *Soil.bulk_ec_dc_tc* (see Figure 8) using the methods described in Section 2.1, 2.2 and 2.4. The main function *Water* follows a hierarchical data fusion that prioritizes predictions based on *Soil.bulk_perm* (in *predict.water_from_perm.py*) if *Soil.frequency_perm* is known. Alternatively, *Soil.water* is calculated based on *Soil.bulk_ec_dc_tc* (in *predict.water_from_ec.py*). The workflows of *predict.water_from_perm.py* and *predict.water_from_ec.py* are similar to those in *predict.bulk_perm.py* and *predict.bulk_ec_dc_tc.py*, respectively.

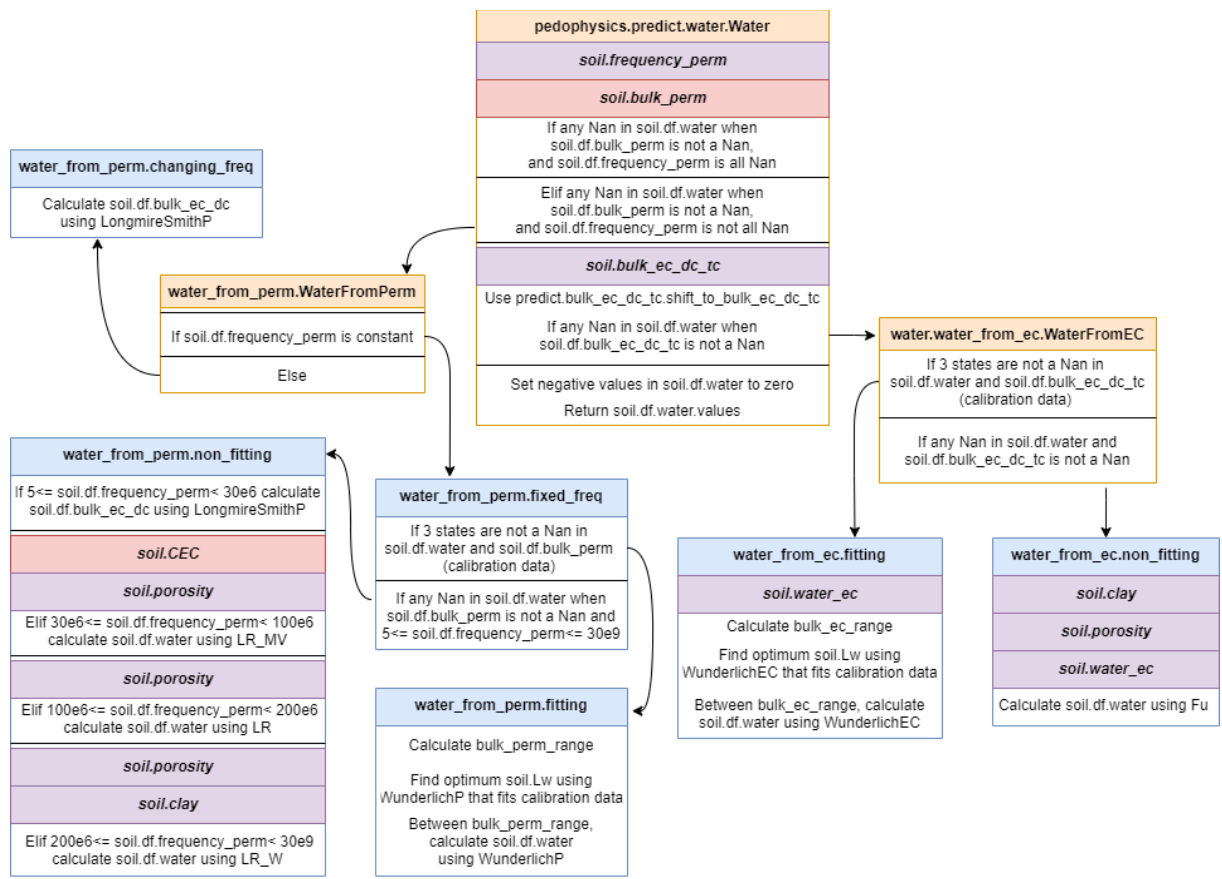

*Figure 8 Prediction of volumetric water content workflow based on Soil.bulk_perm and Soil.bulk_ec_dc_tc. 'Water' is the main function that ties all the other functions together, while light blue headings refer to secondary functions. Each block describes a function, which name is at the heading. Black horizontal lines divide a block into different conditions. Light red blocks refer to mandatory soil attributes for a condition. Purple blocks refer to mandatory soil attributes that can be derived from other ones.*

The function *WaterFromPerm* is used when for any state *Soil.bulk_perm* is given and *Soil.water* is a NaN. This function evaluates if *Soil.frequency_perm* is constant or dynamic. In the first case, the function *fixed_freq* checks for the availability of calibration data, leading to a *fitting* or *non_fitting* function. When *Soil.frequency_perm* is dynamic (thus varying through time or space), *Soil.bulk_ec_dc* is calculated through *changing_freq* using the *LongmireSmithP* function. Afterwards, *Soil.bulk_ec_dc* will be utilized in the *WaterFromEC* function.

Subsequently, the function *shift_to_bulk_ec_dc_tc*, outlined in Figure 7 Pedophysics workflow for the calculation of $\sigma_{dc\,tc}$. Each block describes a function, which name is at the heading. Light yellow heading is used for the main function predict.bulk_ec_dc_tc.BulkECDCTC, while light blue headings refer to secondary functions. Black horizontal lines divide a block into different conditions. Light red blocks refer to mandatory soil attributes for a condition. Purple blocks refer to mandatory soil attributes that can be derived from other ones., tackles the completion of *Soil.bulk_ec_dc_tc*. When this data is available (and *Soil.water* is unknown), the *WaterFromEC* function is applied following a fitting approach using *WunderlichEC,* or a non-fitting approach employing *Fu*.

Finally, predicted *Soil.water* can have negative values in some state because of extrapolating a fitted model to low values of *Soil.bulk_ec_dc_tc* or *Soil.bulk_perm*. In such case, negative values are converted to zero. Then, *Soil.df.water.values* is returned.

**3.2.3 predict soil salinity and pore water electrical conductivity.**

Predicting *Soil.salinity* and *Soil.water_ec* in *Pedophysics* is done by the methods described in Section 2.3 and Figure 9. Pedophysics workflow for calculating salinity and $\sigma_w$. Each block describes a function, the name of which is presented in the block headers. Orange headers are used for main functions (predict.water_ec.WaterEC and predict.salinity.Salinity), while light blue headings refer to secondary functions. Black horizontal lines divide a block into different conditions. Light red blocks refer to mandatory soil attributes. Purple blocks refer to mandatory soil attributes that can be derived from other ones.. The attributes *Soil.salinity* and *Soil.water_ec* are converted one from the other using the *SenGoode* function (Equation 10). Calculating *Soil.water_ec* follows fitting approaches when calibration data are available. If the data include *Soil.water* and *Soil.bulk_ec_dc_tc* for at least two states, the Rhoades approach (Equation 5) is used. Conversely, if the data include *Soil.bulk_ec_dc_tc* and *Soil.bulk_perm* (>10) for at least two states, the Hilhorst approach (outlined in Equation 12) is applied. Finally, if some NaN remains in *Soil.water_ec* while *Soil.water* and *Soil.bulk_ec_dc_tc* are not a NaN, a non-fitting approach uses the *Fu* model.

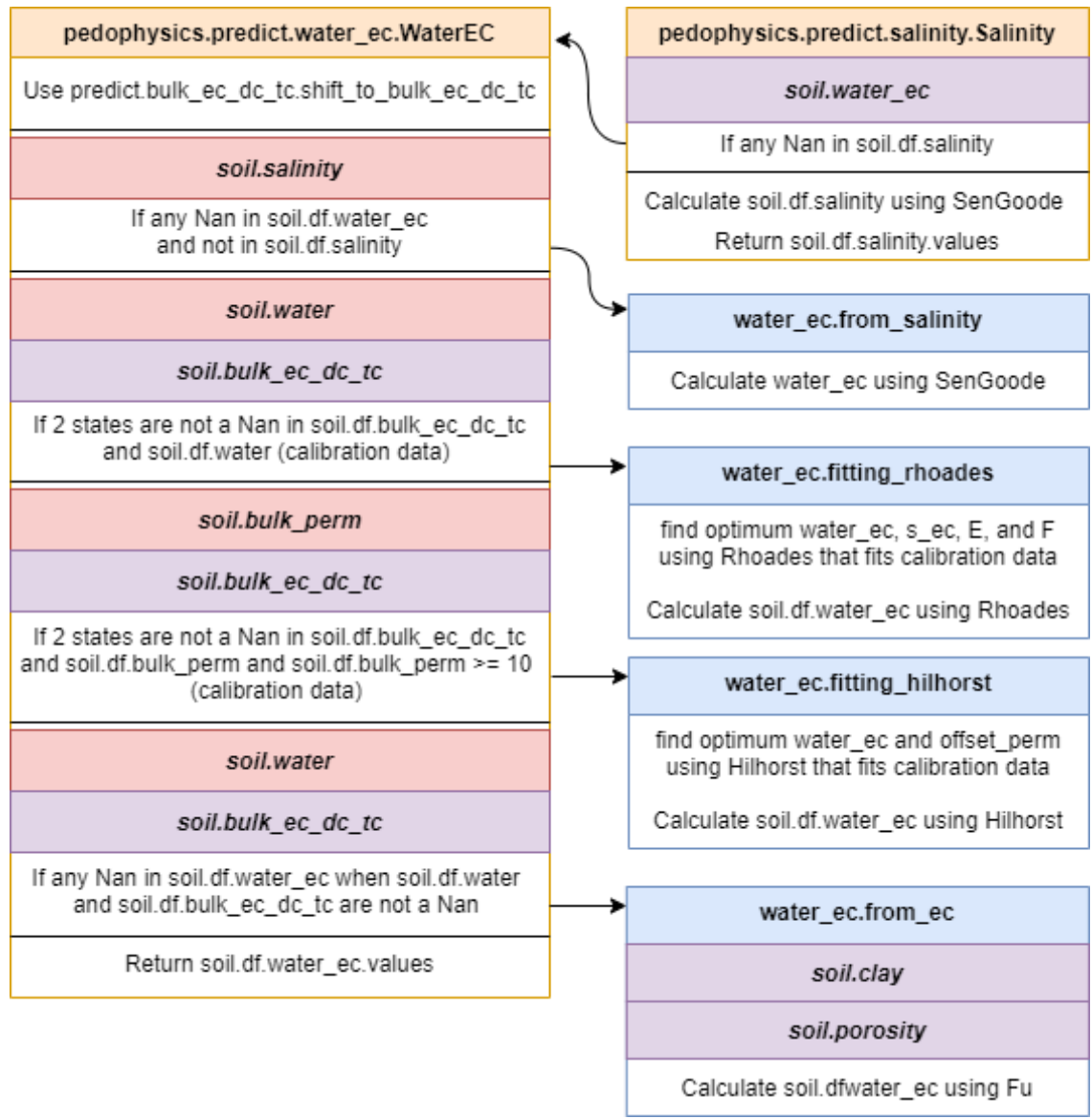

Figure 9. Pedophysics workflow for calculating salinity and $\sigma_w$. Each block describes a function, the name of which is presented in the block headers. Orange headers are used for main functions (predict.water_ec.WaterEC and predict.salinity.Salinity), while light blue headings refer to secondary functions. Black horizontal lines divide a block into different conditions. Light red blocks refer to mandatory soil attributes. Purple blocks refer to mandatory soil attributes that can be derived from other ones.

# 4. Results

This section provides practical coding examples for various scenarios in soil geophysical prospection and pedophysical modelling. For direct access to the codes here reproduced, the reader is referred to the Jupyter notebooks ('Pedophysics/examples' folder) on GitHub.

**4.1 Case of single frequency data acquisition without calibration.**

As a first example, we consider the scenario of an ERT survey whereby a sequence of $\sigma_b$ values are obtained, and for which we aim to predict $\theta$. First, a *Soil* instance named *sample2* is defined by providing attributes (*Soil.clay*, *Soil.porosity*, and *Soil.water_ec*, **Error! Reference source not found.**). The *predict.Water* function is then used to estimate the target attribute *Soil.water* for *sample2*. In this process, *Pedophysics* automatically selects and applies the optimal approach to determine *Soil.water* based on the provided attributes. Information about the applied modeling approach can then be retrieved via the *info* method (*sample2.info.water*). In this case, *info* reveals that a non-fitting approach, implementing the *Fu* function, was used. The metadata of the given attribute *bulk_ec* (*sample2.info.bulk_ec*) details that this is given and not calculated. As a result, a prediction of *Soil.water* is obtained for each soil state, which is printed and plotted (code not shown). A NaN value was given for the state three in *bulk_ec* (i.e., sample2.bulk_ec[3] is a NaN). Consequently, the prediction of *Water* for such state is also a NaN, because it does not meet the mandatory requirements for modelling in *predict.water_from_ec.non_fitting* (this is: *Soil.bulk_ec*, *Soil.clay*, *Soil.porosity*, and *Soil.water_ec*, see Figure 8 Prediction of volumetric water content workflow based on Soil.bulk_perm and Soil.bulk_ec_dc_tc. 'Water' is the main function that ties all the other functions together, while light blue headings refer to secondary functions. Each block describes a function, which name is at the heading. ).

Input:

```
sample2 = Soil( bulk_ec = [0.01, 0.02, 0.025, np.nan, 0.030, 0.040],
                clay = 10,
                porosity = 0.4,
                water_ec = 0.5)

sample2_water = predict.Water(sample2)
print('sample2_water', sample2_water)
print('sample2.info.water', sample2.info.water)
print('sample2.info.bulk_ec', sample2.info.bulk_ec)
```

Output:

```
sample2_water [0.11  0.167 0.19   nan 0.211 0.249]

sample2.info.water
0    nan--> Calculated using Fu function (reported R2=0.98) in predict.water_from_ec.non_fitting
1    nan--> Calculated using Fu function (reported R2=0.98) in predict.water_from_ec.non_fitting
2    nan--> Calculated using Fu function (reported R2=0.98) in predict.water_from_ec.non_fitting
3    nan--> Calculated using Fu function (reported R2=0.98) in predict.water_from_ec.non_fitting
4    nan--> Calculated using Fu function (reported R2=0.98) in predict.water_from_ec.non_fitting
5    nan--> Calculated using Fu function (reported R2=0.98) in predict.water_from_ec.non_fitting

sample2.info.bulk_ec
0    Value given by the user
1    Value given by the user
2    Value given by the user
3                        NaN
4    Value given by the user
5    Value given by the user
```

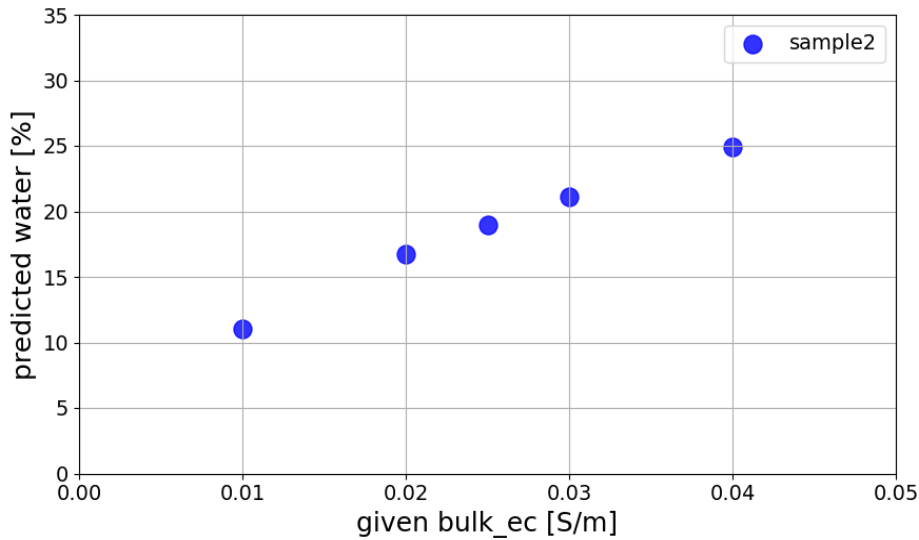

*Figure 10 Practical example of a soil with dynamic Soil.bulk_ec, and Soil.water as a target. The Soil instance ´sample2´ has four defined attributes (Soil.bulk_ec, Soil.clay, Soil.porosity, and Soil.water_ec). The prediction of Soil.water is possible because all the required attributes are provided, except for the state 3. The plot (its code is not reproduced here) shows the variation of the given Soil.bulk_ec versus the predicted Soil.water.*

Similarly, to simulate that every $\sigma_b$ measurement has been obtained in a slightly different soil texture, *Soil.clay* is provided for the six soil states (**Error! Reference source not found.**). Then, the prediction of *Soil.water* (in red) considers a dynamic *Soil.clay*.

Input:
```
sample3 = Soil(bulk_ec = [0.01, 0.02, 0.025, np.nan, 0.030, 0.040],
               clay = [11, 8, 12, 10, 15, 7],
               porosity = 0.4,
               water_ec = 0.5)

sample3_water = predict.Water(sample3)
print('sample3_water', sample3_water)
```

Output:
```
sample3_water [0.107 0.172 0.185   nan 0.197 0.257]
```

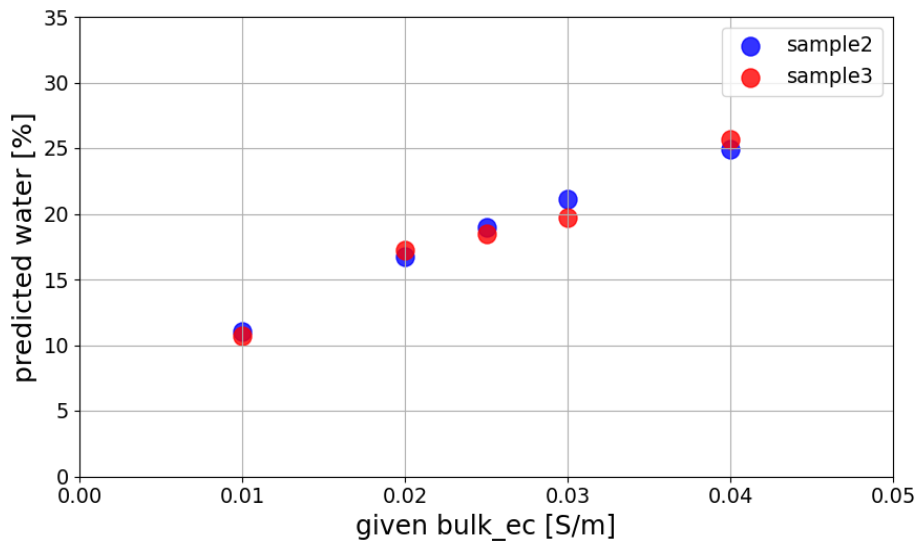

*Figure 101 Practical example of a soil with dynamic Soil.bulk_ec and Soil.clay, with Soil.water as a target. The Soil instance ´sample3´ has four defined attributes (Soil.bulk_ec, Soil.clay, Soil.porosity, and Soil.water_ec). The prediction of Soil.water is possible because all the required attributes are provided, except for the state 3. The plot (its code is not reproduced here) shows the variation of the given Soil.bulk_ec versus the predicted Soil.water.*

### 4.2 Integrating calibration data.

Adding to the last example, the standard use of pedophysical modelling can be optimized by implementing soil-specific solutions. This is normally done by performing a soil-calibration, either in field or laboratory conditions. The present section shows how the decision workflow of *pedophysics* prioritizes a soil-specific solution over a standard one.

Consider a forward modelling simulation where the effect of $\theta$ on $\varepsilon_b$ (measured using GPR) is investigated. Hereby, *Soil* instance *sample4* would be defined with a dynamic *Soil.water*, as shown in Figure 11, allowing the prediction of *Soil.bulk_perm*. The metadata in *sample4.info.bulk_perm* shows it was calculated using the *LR_W* function, which is a suitable PPM for the EM frequencies at which GPR instruments typically operate. To show how *pedophysics* automatically completes missing information if possible, the attribute *Soil.water_perm* is printed. Despite the unknown *Soil.water_perm*, it is set to 80, which allows populating the *LR_W* function (details in *sample4.info.water_perm*). Additionally, because *Soil.porosity* is a mandatory attribute for *LR_W* use, this is automatically derived from the given *Soil.bulk_density*.

**Input:**

```
sample4 = Soil( water = [0.01, 0.1, 0.125, 0.025, 0.05, 0.18, 0.22],
                clay = 20,
                bulk_density = 1.35,
                temperature = 15+273.15,
                instrument = 'GPR')

sample4_bulk_perm = predict.BulkPerm(sample4)
print('sample4_bulk_perm', sample4_bulk_perm)
print('sample4.info.bulk_perm', sample4.info.bulk_perm)
print('sample4.water_perm', sample4.water_perm)
print('sample4.df.water_perm.values', sample4.df.water_perm.values)
print('sample4.df.porosity.values', sample4.df.porosity.values)
print('sample4.info.porosity[0]', sample4.info.porosity[0])
```

**Output:**

```
sample4_bulk_perm [ 2.796  6.36   7.537  3.312  4.244 10.385 12.669]
sample4.info.bulk_perm
0    nan--> Calculated using LR_W function in predict.bulk_perm.non_fitting
1    nan--> Calculated using LR_W function in predict.bulk_perm.non_fitting
2    nan--> Calculated using LR_W function in predict.bulk_perm.non_fitting
3    nan--> Calculated using LR_W function in predict.bulk_perm.non_fitting
4    nan--> Calculated using LR_W function in predict.bulk_perm.non_fitting
5    nan--> Calculated using LR_W function in predict.bulk_perm.non_fitting
6    nan--> Calculated using LR_W function in predict.bulk_perm.non_fitting
sample4.water_perm [nan nan nan nan nan nan nan]
sample4.df.water_perm.values [80 80 80 80 80 80 80]
sample4.df.porosity.values [0.491 0.491 0.491 0.491 0.491 0.491 0.491]
sample4.info.porosity[0] Calculated based on bulk density
```

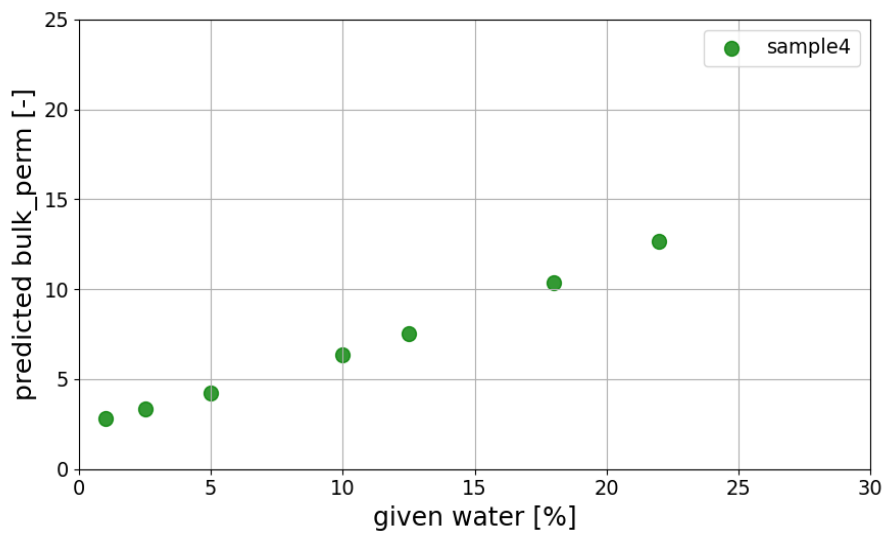

Figure 112 Practical example of a soil with dynamic Soil.water, and Soil.bulk_perm as a target. The Soil instance ´sample4´ has five defined attributes (Soil.water, Soil.clay, Soil.bulk_density, Soil.temperature, and Soil.instrument). The prediction of Soil.bulk_perm is possible because all the required attributes are provided. The plot (its code is not reproduced here) shows the variation of the given Soil.water versus the predicted Soil.bulk_perm.

Calibration data added to *sample4* (*sample4b*, Figure 12) includes three soil states with given *Soil.water* and *Soil.bulk_perm* values (indicated by black triangles in Figure 12 plot). The workflow of *pedophysics* automatically decides for a solution through fitting of the calibration data (see Figure 6. Pedophysics workflow for the calculation of $\varepsilon_b$. Each block describes a function, which name is at the heading. Light yellow heading is used for the main function predict.bulk_perm.BulkPerm, while light blue headings refer to secondary functions. Black horizontal lines divide a block into different conditions. Light red blocks refer to mandatory soil attributes for a condition. Purple blocks refer to mandatory soil attributes that can be derived from other ones.), plotted in red. This allows to obtain a precise and specific solution that can be extrapolated to different values of *Soil.water*. A closer look into *.info* shows that the fitting approach used the *WunderlichP* function, which provided a good fit of the calibration data.

**Input:**

```python
sample4b = Soil(water = [0.17, 0.03, 0.07, 0.01, 0.1, 0.125, 0.025, 0.05, 0.18, 0.22],
                bulk_perm = [12, 4.5, 7],
                clay = 20,
                bulk_density = 1.35,
                temperature = 15+273.15,
                instrument = 'GPR')

sample4b_bulk_perm = predict.BulkPerm(sample4b)
print('sample4b_bulk_perm', sample4b_bulk_perm)
print('sample4b.info.bulk_perm', sample4b.info.bulk_perm)
```

**Output:**

```
sample4b_bulk_perm [12.     4.5    7.     3.669  8.003  9.434  4.284  5.413 12.829 15.468]
sample4b.info.bulk_perm
0      Value given by the user
1      Value given by the user
2      Value given by the user
3      nan--> Calculated by fitting (R2=0.986) WunderlichP function in predict.bulk_perm.fitting, for
soil.water values between[0, 0.24]
4      nan--> Calculated by fitting (R2=0.986) WunderlichP function in predict.bulk_perm.fitting, for
soil.water values between[0, 0.24]
5      nan--> Calculated by fitting (R2=0.986) WunderlichP function in predict.bulk_perm.fitting, for
soil.water values between[0, 0.24]
6      nan--> Calculated by fitting (R2=0.986) WunderlichP function in predict.bulk_perm.fitting, for
soil.water values between[0, 0.24]
.
.
.
```

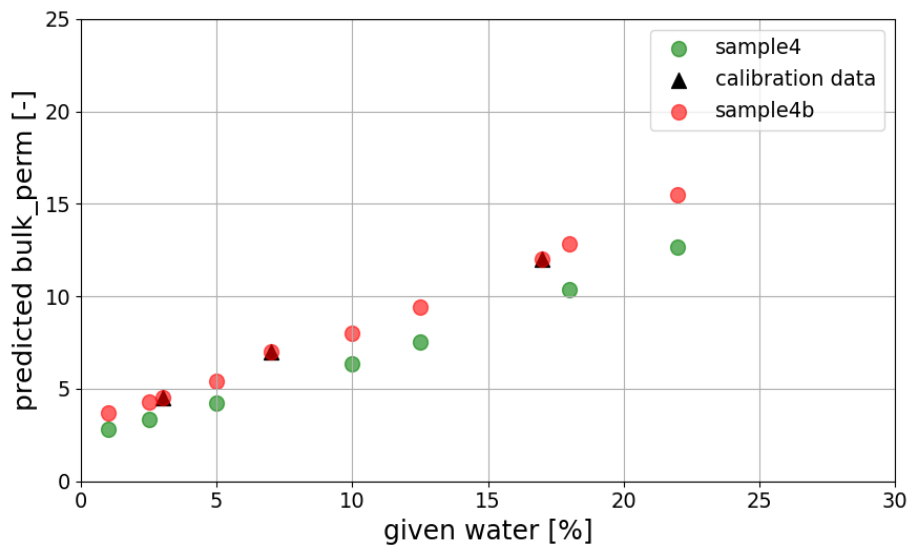

*Figure 12 Practical example of a soil with dynamic Soil.water, and Soil.bulk_perm as a target including calibration data. The Soil instance ´sample4b´ has six defined attributes (Soil.water, Soil.clay, Soil.bulk_perm, Soil.bulk_density, Soil.temperature, and Soil.instrument). The prediction of Soil.bulk_perm is possible because all the required attributes are provided. The plot shows the variation of the given Soil.water versus the predicted Soil.bulk_perm (orange dots), black triangles show the given calibration data, and the previous solution of sample4 remains in green.*

## 4.3 dielectric dispersion data

This section shows a practical case of geophysical data fusion by considering a field survey using ERT and EMI techniques (both work at different EM frequencies), with $\theta$ as the target.

This scenario is implemented in *pedophysics* (Figure 13), where the survey data is given in *sample5.bulk_ec*, and *sample5.frequency_ec* specifies the EM frequency of acquisition. The first five states correspond to ERT data (in DC regime), and the rest of states to EMI data (observed at 50 kHz). The target *sample5.water* was calculated using *Fu* in a non-fitting approach. Because *Soil.bulk_ec* is given but not *Soil.bulk_ec_dc_tc*, the prints of *.info* detail the conversions (using as example the state 6). The result is plotted and shows different trends for the different techniques, showing that considering EM frequencies in pedophysical modelling is relevant for $\theta$ prediction.

Input:

```
sample5 = Soil(
    bulk_ec = [0.009, 0.0125, 0.005, 0.008, 0.001, 0.0085, 0.012, 0.0045, 0.006, 0.002],
    frequency_ec = [0, 0, 0, 0, 0, 50e3, 50e3, 50e3, 50e3, 50e3],
    texture = 'Sand', bulk_density= 1.5, water_ec = 0.3)

sample5_water = predict.Water(sample5)
print('sample5_water', sample5_water)
print('sample5.info.water', sample5.info.water)
print('sample5.info.bulk_ec_dc_tc[6]', sample5.info.bulk_ec_dc_tc[6])
print('sample5.info.bulk_ec_dc[6]', sample5.info.bulk_ec_dc[6])
```

Output:

```
sample5_water [0.15  0.182 0.109 0.142 0.042 0.136 0.167 0.092 0.11  0.054]
sample5.info.water
0    nan--> Calculated using Fu function (reported R2=0.98) in predict.water_from_ec.non_fitting
1    nan--> Calculated using Fu function (reported R2=0.98) in predict.water_from_ec.non_fitting
2    nan--> Calculated using Fu function (reported R2=0.98) in predict.water_from_ec.non_fitting
3    nan--> Calculated using Fu function (reported R2=0.98) in predict.water_from_ec.non_fitting
4    nan--> Calculated using Fu function (reported R2=0.98) in predict.water_from_ec.non_fitting
.
.
sample5.info.bulk_ec_dc_tc[6]
nan--> Equal to soil.df.bulk_ec_dc in predict.bulk_ec_dc_tc.non_tc_to_tc
sample5.info.bulk_ec_dc[6]
nan--> EM frequency shift from actual to zero Hz using LongmireSmithEC function in
predict.bulk_ec_dc.non_dc_to_dc
```

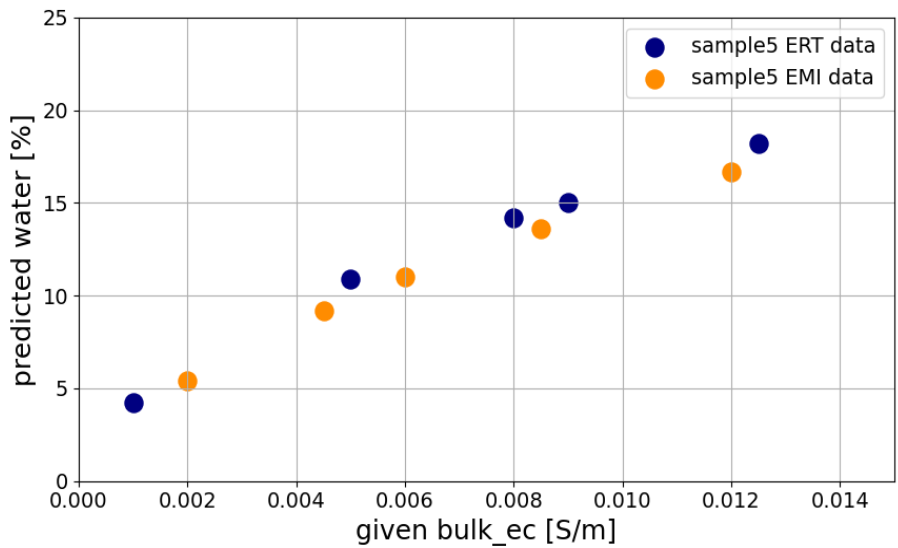

*Figure 13 Practical example of a soil with dynamic Soil.bulk_ec and Soil.frequency_ec, with Soil.water as a target. The Soil instance ´sample5´ has five defined attributes (Soil.bulk_ec, Soil.frequency_ec, Soil.texture, Soil.bulk_density, and Soil.water_ec). The prediction of water is possible because all the required attributes are provided. The plot (its code is not reproduced here) shows the variation of the given Soil.bulk_ec versus the predicted Soil.water for the different EM frequencies.*

Adding to the last example, *sample5b* includes calibration data for the first three soil states (Figure 14); this maybe the case when soil samples for water content are collected in the field along the ERT. Then, the solution for *Soil.water* follows a fitting approach that returned a specific solution (different as that for *sample5*).

**Input:**

```
sample5b = Soil(
    bulk_ec = [0.009, 0.0125, 0.005, 0.008, 0.001, 0.0085, 0.012, 0.0045, 0.006, 0.002],
    water = [0.17, 0.21, 0.115],
    frequency_ec = [0, 0, 0, 0, 0, 50e3, 50e3, 50e3, 50e3, 50e3],
    texture = 'Sand', bulk_density= 1.5)

sample5b_water = predict.Water(sample5b)
print('sample5b_water', sample5b_water)
print('sample5b.info.water', sample5b.info.water)
print('sample5b.df.water_ec.values', sample5b.df.water_ec.values)
print('sample5b.info.water_ec[0]', sample5b.info.water_ec[0])
```

**Output:**

```
sample5b_water [0.17  0.21  0.115 0.156 0.041 0.148 0.189 0.097 0.117 0.056]
sample5b.info.water
0      Value given by the user
1      Value given by the user
2      Value given by the user
3    nan--> Calculated by fitting (R2=0.999) WunderlichEC function in predict.water_from_ec.fit-
ting, for soil.bulk_ec values between: [0.001, 0.016]
4    nan--> Calculated by fitting (R2=0.999) WunderlichEC function in predict.water_from_ec.fit-
ting, for soil.bulk_ec values between: [0.001, 0.016]
5    nan--> Calculated by fitting (R2=0.999) WunderlichEC function in predict.water_from_ec.fit-
ting, for soil.bulk_ec values between: [0.001, 0.016]
.
.
.
sample5b.df.water_ec.values
[0.097859 0.097859 0.097859 0.097859 0.097859 0.097859 0.097859 0.097859 0.097859 0.097859]
sample5b.info.water_ec[0]
nan--> Calculated by fitting (R2 = 0.983) Rhoades function in predict.water_ec.fitting_rhoades
```

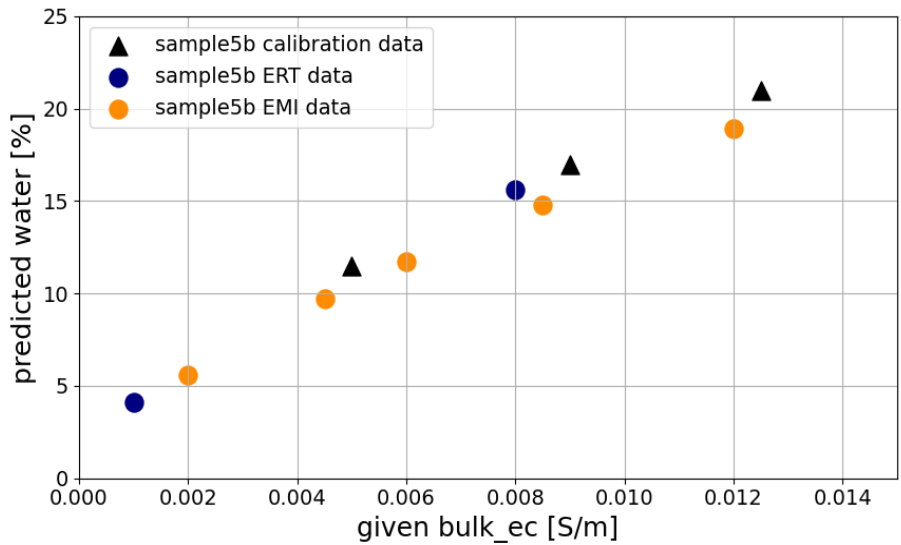

*Figure 14 Practical example of a soil with dynamic Soil.bulk_ec and Soil.frequency_ec, whit Soil.water as a target and including calibration data. The Soil instance ´sample5b´ has five defined attributes (Soil.bulk_ec, Soil.water, Soil.frequency_ec, Soil.texture, and Soil.bulk_density) The prediction of water is possible because all the required attributes are provided. The plot (code not reproduced here) shows the variation of the given Soil.bulk_ec versus the predicted Soil.water for the different EM frequencies, including calibration data (black triangles).*

**4.4 soil salinity**

In this section, two different approaches for calculating *Soil.water_ec* and *Soil.salinity* are presented as examples. The first case represent a soil measured using TDR, where $\sigma_b$ and $\varepsilon_b$ are observed, with a reported $\sigma_w = 0.29$ S/m (data from Hamed et al., 2003, Odarslöv topsoil). The data of such a soil is defined in *sample6,* and the function *predict.WaterEC* is used (**Error! Reference source not found.**), obtaining a similar result as the reported. The metadata (in *sample6.info.water_ec*) shows the fitting accuracy (R2 = 0.997) and the used *Hilhorst* function. Additionally, to reproduce the fitting of the calibration data, *Hilhorst* is called with the *sample6* attributes.

Input:
```
from pedophysics.pedophysical_models.bulk_perm import Hilhorst

sample6 = Soil(bulk_ec = [0.02, 0.03, 0.04, 0.05, 0.06],
               bulk_perm = [11.5, 14.8, 17, 20, 22.7],
               clay=5,
               bulk_density=1.48,
               instrument='TDR')

sample6_water_ec = predict.WaterEC(sample6)
print('sample6_water_ec', sample6_water_ec)
print('sample6.info.water_ec', sample6.info.water_ec)
sample6_fitted = Hilhorst(sample6.df.bulk_ec, sample6.df.water_ec, sample6.df.water_perm, sample6.df.offset_perm)
```
Output:
```
sample6_water_ec [0.289855 0.289855 0.289855 0.289855 0.289855]
sample6.info.water_ec
0    nan--> Calculated by fitting (R2=0.997) Hilhorst function in predict.water_ec.fitting_hilhorst
1    nan--> Calculated by fitting (R2=0.997) Hilhorst function in predict.water_ec.fitting_hilhorst
2    nan--> Calculated by fitting (R2=0.997) Hilhorst function in predict.water_ec.fitting_hilhorst
3    nan--> Calculated by fitting (R2=0.997) Hilhorst function in predict.water_ec.fitting_hilhorst
4    nan--> Calculated by fitting (R2=0.997) Hilhorst function in predict.water_ec.fitting_hilhorst
Name: water_ec, dtype: object
```

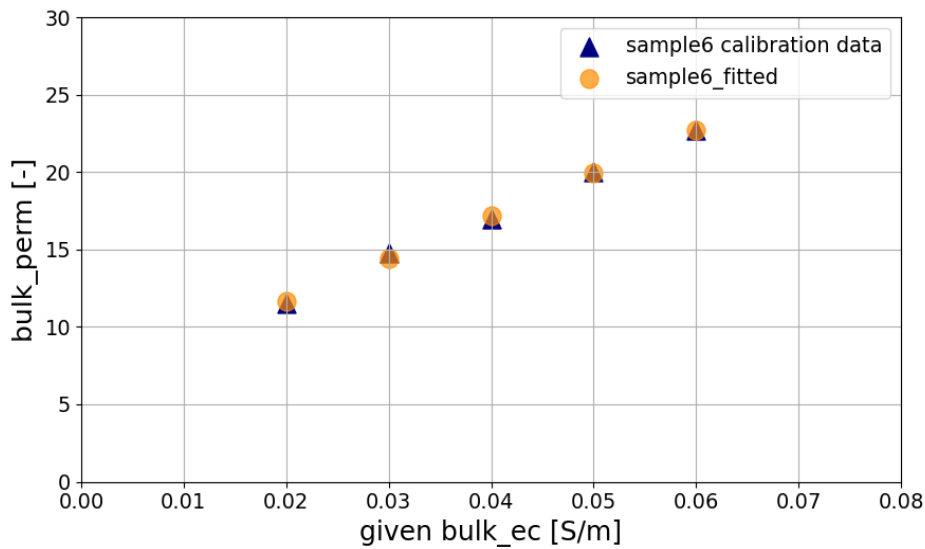

*Figure 15 Practical example of a soil with calibration data of Soil.bulk_ec and Soil.bulk_perm, with Soil.water_ec as a target. The Soil instance ´sample6´ has five defined attributes (Soil.bulk_ec, Soil.bulk_perm, Soil.clay, Soil.bulk_density, and Soil.instrument). The prediction of Soil.water_ec is possible because all the required attributes are provided. The plot (its code is not reproduced here) shows the variation of the given Soil.bulk_ec versus Soil.bulk_perm, and the fitting of these calibration data.*

A similar case is the example of *sample5b* (Figure 14)*,* where based on the calibration data of *Soil.water* and *Soil.bulk_ec*, *Soil.water_ec* is automatically calculated (see the print of *sample5b.water_ec* and its metadata) using *Rhoades* (Equation 5). This is because *Soil.water_ec* is a requirement to calculate the target attribute *Soil.water* using *WunderlichEC*. Further details and an additional example using *Rhoades* with data from Brovelli and Cassiani (2011) can be seen in the code implementation of this section (Examples.ipynb).

# 5 limitations

The primary limitations of *Pedophysics* stem from the performance of the implemented PMs and PTFs. While these models have been validated extensively across a wide range of soil conditions, they may fall short in representing extreme scenarios outside the tested soil characteristics ranges.

Pedophysical modelling, in general, is limited by gaps in PM developments for specific conditions. In particular, the prediction of $\varepsilon_b$ in the frequency range of 100 MHz to 200 MHz remains uncertain primarily due to the Maxwell-Wagner effect, and its complex relationship with soil texture. Moreover, *LongmireSmithP* catches the $\varepsilon_b$ response up to approximately 200 MHz, while for higher frequencies the model is just validated for sandy samples. As a result, *Pedophysics* is constrained according to these limitations for $\varepsilon_b$ dielectric dispersion. Additionally, the implemented *Fu* model (Equation 8) shows limitations in its application to soils with clay content higher than 33%.

# 6. Conclusions and further developments

Optimal published PMs and PTFs were selected and implemented in the *Pedophysics* python package. The automated workflow of *Pedophysics* enables deploying appropriate PMs and PTFs in different modelling approaches, based on user input. When attributes required for populating PMs or PTFs in predicting target properties are unavailable, these are automatically calculated or assumed where feasible. This approach enables users to obtain optimal solutions with minimum of information about the studied soil.

*Pedophysics* effectively manages complex data fusion scenarios, including soil properties changing over time or space, integrating dielectric dispersion of geophysical properties, and cases with available calibration data. This way, users can rely on *Pedophysics* to implement specific knowledge about pedophysical modeling. Ultimately, the aim of *Pedophysics* is to contribute to the facilitation of worldwide access to soil geophysical solutions.

The open-source nature of code and structure of the software facilitate easy maintenance and the addition of new features by interested researchers in the community. Pedophysical modelling could be enhanced by integrating additional features into *Pedophysics*. These improvements could include incorporating both apparent and imaginary components of electrical conductivity and dielectric permittivity, and comprehensive pedotransfer functions for $CEC$ tested with worldwide soils. Furthermore, additions such as a fourth soil phase for estimating contaminants could be incorporated, alongside capability to predict matric potential and estimate soil properties from seismic and magnetic geophysical techniques.

## Acknowledgements

## Formatting of funding sources

This work was supported by the Ghent University Special Research Fund (BOF), grant number 01IO4120.

## Computer Code Availability

Pedophysics 1.0. Developer: Gaston Mendoza Veirana, gaston.mendozaveirana@ughent.be. Available since 28th of February 2024. Hardware required: Hardware minimum requirements: i3, two nucleus processor, 8 GB ram memory. Software required: see requirements.txt. Program language: Python . Source code: https://github.com/orbit-ugent/Pedophysics. Tests available for Pytest library. GNU General Public License v3.0

## Authorship Statement.

**Gaston Mendoza Veirana:** Conceptualization, Data curation, Formal analysis, Investigation, Methodology, Project administration, Software, Supervision, Validation, Visualization, Writing – original draft, Writing – review & editing

**Philippe De Smedt:** Conceptualization, Funding acquisition, Project administration, Resources, Supervision, Writing – review & editing

**Jeroen Verhegge:** Funding acquisition, Resources, Supervision, Writing – review & editing

**Wim Cornelis:** Funding acquisition, Resources, Writing – review & editing

**Declaration of generative AI and AI-assisted technologies in the writing process**

During the preparation of this work the author(s) used GPT-4 OpenAI to improve text readability. After using this tool/service, the author(s) reviewed and edited the content as needed and take(s) full responsibility for the content of the publication.

## References

Alipio, R., & Visacro, S. (2014). Modeling the Frequency Dependence of Electrical Parameters of Soil. *IEEE Transactions on Electromagnetic Compatibility*, *56*(5), 1163–1171. https://doi.org/10.1109/TEMC.2014.2313977

Amente, G., Baker, J. M., & Reece, C. F. (2000). Estimation of Soil Solution Electrical Conductivity from Bulk Soil Electrical Conductivity in Sandy Soils. *Soil Science Society of America Journal*, *64*(6), 1931–1939. https://doi.org/10.2136/sssaj2000.6461931x

Archie, G. E. (1942). The Electrical Resistivity Log as an Aid in Determining Some Reservoir Characteristics. *Transactions of the AIME*, *146*(01), 54–62. https://doi.org/10.2118/942054-G

Bañón, S., Álvarez, S., Bañón, D., Ortuño, M. F., & Sánchez-Blanco, M. J. (2021). Assessment of soil salinity indexes using electrical conductivity sensors. *Scientia Horticulturae*, *285*, 110171. https://doi.org/10.1016/j.scienta.2021.110171

Bennett, D. L., George, R. J., & Whitfield, B. (2000). The use of ground EM systems to accurately assess salt store and help define land management options for salinity management. *Exploration Geophysics*, *31*(1–2), 249–254. https://doi.org/10.1071/EG00249

Bigelow, R. C., & Eberle, W. R. (1983). *Empirical predictive curves for resistivity and dielectric constant of earth materials; 100 Hz to 100 MHz* (Report 83–911; Open-File Report). USGS Publications Warehouse. https://doi.org/10.3133/ofr83911

Brovelli, A., & G. Cassiani. (2011). Combined estimation of effective electrical conductivity and permittivity for soil monitoring. *WATER RESOURCES RESEARCH*. https://doi.org/10.1029/2011WR010487

Cavka, D., Mora, N., & Rachidi, F. (2014). A Comparison of Frequency-Dependent Soil Models: Application to the Analysis of Grounding Systems. *IEEE Transactions on Electromagnetic Compatibility*, *56*(1), 177–187. https://doi.org/10.1109/TEMC.2013.2271913

Chen, Y., & Or, D. (2006a). Effects of Maxwell-Wagner polarization on soil complex dielectric permittivity under variable temperature and electrical conductivity: MAXWELL-WAGNER

EFFECT ON SOIL PERMITTIVITY. *Water Resources Research*, *42*(6).

https://doi.org/10.1029/2005WR004590

Chen, Y., & Or, D. (2006b). Geometrical factors and interfacial processes affecting complex dielectric permittivity of partially saturated porous media: COMPLEX DIELECTRIC PERMITTIVITY OF POROUS MEDIA. *Water Resources Research*, *42*(6). https://doi.org/10.1029/2005WR004744

Ciampalini, A., André, F., Garfagnoli, F., Grandjean, G., Lambot, S., Chiarantini, L., & Moretti, S. (2015). Improved estimation of soil clay content by the fusion of remote hyperspectral and proximal geophysical sensing. *Journal of Applied Geophysics*, *116*, 135–145. https://doi.org/10.1016/j.jappgeo.2015.03.009

Corwin, D. L., & Lesch, S. M. (2005). Characterizing soil spatial variability with apparent soil electrical conductivity: Part II. Case study. *Computers and Electronics in Agriculture*, *46*(1–3), 135–152.

Corwin, D. L., & Plant, R. E. (2005). Applications of apparent soil electrical conductivity in precision agriculture. *Computers and Electronics in Agriculture*, *46*(1–3), 1–10. https://doi.org/10.1016/j.compag.2004.10.004

Corwin, D. L., & Yemoto, K. (2020). Salinity: Electrical conductivity and total dissolved solids. *Soil Science Society of America Journal*, *84*(5), 1442–1461. https://doi.org/10.1002/saj2.20154

Dobson, M., Ulaby, F., Hallikainen, M., & El-rayes, M. (1985). Microwave Dielectric Behavior of Wet Soil-Part II: Dielectric Mixing Models. *IEEE Transactions on Geoscience and Remote Sensing*, *GE-23*(1), 35–46. https://doi.org/10.1109/TGRS.1985.289498

Doussan, C., & Ruy, S. (2009). Prediction of unsaturated soil hydraulic conductivity with electrical conductivity: SOIL HYDRAULIC-ELECTRICAL CONDUCTIVITY. *Water Resources Research*, *45*(10). https://doi.org/10.1029/2008WR007309

Fu, R. R., Maher, B. A., Nie, J., Gao, P., Berndt, T., Folsom, E., & Cavanaugh, T. (2022). *Pinpointing the mechanism of magnetic enhancement in modern soils using high-resolution magnetic field imaging* [Preprint]. Preprints. https://doi.org/10.22541/essoar.167160976.60846594/v1

Glover, P. W. J. (2015). Geophysical Properties of the Near Surface Earth: Electrical Properties. In *Treatise on Geophysics* (pp. 89–137). Elsevier. https://doi.org/10.1016/B978-0-444-53802-4.00189-5

Glover, P. W. J., Hole, M. J., & Pous, J. (2000). A modified Archie's law for two conducting phases. *Earth and Planetary Science Letters*, *180*(3), 369–383. https://doi.org/10.1016/S0012-821X(00)00168-0

González-Teruel, J. D., Jones, S. B., Soto-Valles, F., Torres-Sánchez, R., Lebron, I., Friedman, S. P., & Robinson, D. A. (2020). Dielectric Spectroscopy and Application of Mixing Models Describing Dielectric Dispersion in Clay Minerals and Clayey Soils. *Sensors*, *20*(22), 6678. https://doi.org/10.3390/s20226678

Hallikainen, M., Ulaby, F., Dobson, M., El-rayes, M., & Wu, L. (1985). Microwave Dielectric Behavior of Wet Soil-Part 1: Empirical Models and Experimental Observations. *IEEE Transactions on Geoscience and Remote Sensing*, *GE-23*(1), 25–34. https://doi.org/10.1109/TGRS.1985.289497

Hamed, Y., Persson, M., & Berndtsson, R. (2003). Soil Solution Electrical Conductivity Measurements Using Different Dielectric Techniques. *Soil Science Society of America Journal*, *67*(4), 1071–1078. https://doi.org/10.2136/sssaj2003.1071

Hanssens, D., Delefortrie, S., De Pue, J., Van Meirvenne, M., & De Smedt, P. (2019). Practical aspects of frequency domain electromagnetic forward and sensitivity modelling of a magnetic dipole in a multi-layered half-space. *IEEE Geoscience and Remote Sensing Magazine*, *7*(1), 74–85. https://doi.org/10.1109/MGRS.2018.2881767

He, J., Zeng, R., & Zhang, B. (2013). *Methodology and technology for power system grounding*. IEEE.

Hilhorst, M. A. (2000). A Pore Water Conductivity Sensor. *Soil Science Society of America Journal*, *64*, 1922–1925.

Jones, J. M. Blonquist, Jr., D. A. Robinson, V. Philip Rasmussen, and D. Or. (2005). Standardizing Characterization of Electromagnetic Water Content Sensors: Part 1. Methodology. *Vadose Zone Journal*, *4*, :1048-1058. https://doi.org/doi:10.2136/vzj2004.0140

Lichtenecker, & Rother. (1931). *"Die Herleitung des logarithmischen Mischungs-gesetzes aus allegemeinen Prinzipien der stationaren Stromung*.

Linford, N. (2006). The application of geophysical methods to archaeological prospection. *Reports on Progress in Physics*, *69*(7), 2205–2257. https://doi.org/10.1088/0034-4885/69/7/R04

Longmire, Conrad L. Smith, Ken S. (n.d.). *A Universal Impedance for Soils*. Topical rept.

Ma, R., McBratney, A., Whelan, B., Minasny, B., & Short, M. (2010). Comparing temperature correction models for soil electrical conductivity measurement. *Precision Agriculture*.

Malicki, M. A., & Walczak, R. T. (1999). Evaluating soil salinity status from bulk electrical conductivity and permittivity: Evaluating soil salinity. *European Journal of Soil Science*, *50*(3), 505–514. https://doi.org/10.1046/j.1365-2389.1999.00245.x

Malmberg, C. G., & Maryott, A. A. (1956). Dielectric constant of water from 0 to 100 C. *Journal of Research of the National Bureau of Standards*, *56*(1), 1. https://doi.org/10.6028/jres.056.001

Mendoza Veirana, G., Verhegge, J., Cornelis, W., & De Smedt, P. (2023). Soil dielectric permittivity modelling for 50 MHz instrumentation. *Geoderma*, *438*, 116624. https://doi.org/10.1016/j.geoderma.2023.116624

Persson, M. (2002). Evaluating the linear dielectric constant-electrical conductivity model using time-domain reflectometry. *Hydrological Sciences Journal*, *47*(2), 269–277. https://doi.org/10.1080/02626660209492929

Revil, A., Karaoulis, M., Johnson, T., & Kemna, A. (2012). Review: Some low-frequency electrical methods for subsurface characterization and monitoring in hydrogeology. *Hydrogeology Journal*, *20*(4), 617–658. https://doi.org/10.1007/s10040-011-0819-x

Revil, A., Schwaeger, H., Cathles, L. M., & Manhardt, P. D. (1999). Streaming potential in porous media: 2. Theory and application to geothermal systems. *Journal of Geophysical Research: Solid Earth*, *104*(B9), 20033–20048. https://doi.org/10.1029/1999JB900090

Rhoades, J. D., Raats, P. A. C., & Prather, R. J. (1976). Effects of Liquid-phase Electrical Conductivity, Water Content, and Surface Conductivity on Bulk Soil Electrical Conductivity. *Soil Sci. Soc. Am. J.*, *40*(5), 651–655. https://doi.org/10.2136/sssaj1976.03615995004000050017x

Romero-Ruiz, A., Linde, N., Keller, T., & Or, D. (2018). A Review of Geophysical Methods for Soil Structure Characterization: Geophysics and soil structure. *Reviews of Geophysics*, *56*(4), 672–697. https://doi.org/10.1029/2018RG000611

Roth, K., Schulin, R., Flühler, H., & Attinger, W. (1990). Calibration of time domain reflectometry for water content measurement using a composite dielectric approach. *Water Resources Research*, *26*(10), 2267–2273. https://doi.org/10.1029/WR026i010p02267

Schjønning, P., McBride, R. A., Keller, T., & Obour, P. B. (2017). Predicting soil particle density from clay and soil organic matter contents. *Geoderma*, *286*, 83–87. https://doi.org/10.1016/j.geoderma.2016.10.020

Schwartz, R. C., Evett, S. R., & Bell, J. M. (2009). Complex Permittivity Model for Time Domain Reflectometry Soil Water Content Sensing: II. Calibration. *Soil Science Society of America Journal*, *73*(3), 898–909. https://doi.org/10.2136/sssaj2008.0195

Sen, P. N., & Goode, P. A. (1992). Influence of temperature on electrical conductivity on shaly sands. *GEOPHYSICS*, *57*(1), 89–96. https://doi.org/10.1190/1.1443191

Sheets, K. R., & Hendrickx, J. M. H. (1995). Noninvasive Soil Water Content Measurement Using Electromagnetic Induction. *Water Resour. Res.*, *31*(10), 2401–2409. https://doi.org/10.1029/95wr01949

Tabbagh, Seger, & Cousin. (2023). Using proximal electromagnetic/electrical resistivity (ER)/electrical impedance spectroscopy sensors to assess soil health and water status. In *Advances in sensor technology for sustainable crop production*. ⟨hal-03839019⟩

Topp, G. C., Hayhoe, H. N., & Watt, M. (1996). Point specific measurement and monitoring of soil water content with an emphasis on TDR. *Canadian Journal of Soil Science*, *76*(3), 307–316. https://doi.org/10.4141/cjss96-037

van Dam, R., Borchers, B., & Hendrickx, J. (2005). *Methods for prediction of soil dielectric properties: A review* (Vol. 5794). SPIE.

Verhegge, J., Mendoza Veirana, G., Cornelis, W., Crombé, P., Grison, H., De Kort, J.-W., Rensink, E., & De Smedt, P. (2023). Developing a geophysical framework for Neolithic land-use studies: In situ monitoring and forward modelling of electrical properties at "Valther-Tweeling" (NL). In T. Wunderlich, H. Hadler, & R. Blankenfeldt (Eds.), *Advances in On- and Offshore Archaeological Prospection* (pp. 399–403). Universitätsverlag Kiel | Kiel University Publishing. https://doi.org/10.38072/978-3-928794-83-1/p81

Waxman, M. H., & Smits, L. J. M. (1968). Electrical Conductivities in Oil-Bearing Shaly Sands. *Society of Petroleum Engineers Journal*, *8*(02), 107–122. https://doi.org/10.2118/1863-A

Wunderlich, T., Petersen, H., Hagrey, S. A. al, & Rabbel, W. (2013). Pedophysical Models for Resistivity and Permittivity of Partially Water-Saturated Soils. *Vadose Zone Journal*, *12*(4), vzj2013.01.0023. https://doi.org/10.2136/vzj2013.01.0023

Zhou, M., Wang, J., Cai, L., Fan, Y., & Zheng, Z. (2015). Laboratory Investigations on Factors Affecting Soil Electrical Resistivity and the Measurement. *IEEE Transactions on Industry Applications*, *51*(6), 5358–5365. https://doi.org/10.1109/TIA.2015.2465931